\documentclass[aps,pra,superscriptaddress,epsfigure,twocolumn,nofootinbib]{revtex4-1}
\usepackage{graphicx}
\usepackage{amsmath}    
\usepackage{latexsym}
\usepackage{amsfonts}   
\usepackage{amssymb}
\usepackage{array}      
\usepackage{epsfig}
\usepackage{txfonts}
\usepackage{hyperref}
\usepackage{color}
\usepackage{braket}
\usepackage{dsfont} 
\usepackage{float}
\usepackage{mathrsfs}
\usepackage{mathtools}

\newcommand\idop{\mathds{1}}

\begin{document}
%
\title{Quantum machines powered by correlated baths}
\author{Gabriele De Chiara}

\affiliation{Centre  for  Theoretical  Atomic,  Molecular  and  Optical  Physics, Queen's  University  Belfast,  Belfast  BT7 1NN,  United  Kingdom}
\affiliation{Laboratoire Charles Coulomb (L2C), UMR 5221 CNRS-Universit\'e de Montpellier, F-34095 Montpellier, France}
\author{Mauro Antezza}
\affiliation{Laboratoire Charles Coulomb (L2C), UMR 5221 CNRS-Universit\'e de Montpellier, F-34095 Montpellier, France}
\affiliation{Institut Universitaire de France, 1 rue Descartes, F-75231 Paris Cedex 05, France}

\begin{abstract}
We consider thermal machines powered by locally equilibrium reservoirs that share classical or quantum correlations.
The reservoirs are modelled by the so-called collisional model or repeated interactions  model. In our framework, two reservoir particles, initially prepared in a thermal  state, are correlated through a unitary transformation and afterwards interact locally with the two  quantum subsystems which form the working fluid. For a particular class of unitaries, we  show how the  transformation applied  to the reservoir particles affects the amount of heat transferred and the work produced. We then compute the distribution of heat  and work  when the unitary  is chosen randomly,  proving that the total swap transformation is the optimal one. Finally, we analyse the performance of the  machines in terms of classical and  quantum correlations established among the microscopic constituents of the  machine.
\end{abstract}

\maketitle

\section{Introduction}
The interest in thermal machines powered by quantum working media has recently surged thanks to the technological advancement in the realisation and control of individual quanta \cite{ThermoBook,XuerebReview,GooldReview,AndersReview,Mitchison2019,deffner2019quantum}. This tremendous progress has led to the first realisations of quantum engines and thermal devices \cite{RossnagelScience2016,ronzani2018tunable,Maslennikov19,LindenfelsPRL2019,KlatzowPRL2019,PetersonPRL2019,gluza2020}. 

The theoretical modelling of such devices usually involves the system  in contact with equilibrium uncorrelated baths at different temperatures. However, some papers have generalised this picture to non equilibrium reservoirs~\cite{LeggioPRA2015,LeggioPRE2016,ReidEPL2017,AssisPRL2019,cherubim2019,PezzuttoQST2019,CarolloPRL2020,AncheytaQST2019}, including the case of the Otto engine in contact with squeezed reservoirs~\cite{RossnagelPRL2014,Manzano2016,AgarwallaPRB2017,singh2020performance}, which can lead to efficiencies and performances beyond the Otto and Carnot limit. This conclusion, obviously, does not take into account the cost of maintaining a non equilibrium reservoir which is then considered as a free resource but shows how to best employ these resources (see also \cite{niedenzu2018quantum}).
Other works have considered thermal devices coupled to spatially separated reservoirs which share correlations, classical or quantum~\cite{DoyeuxPRE2016,TurkpenceEPL2017,KarimiPRB2017,HewgillPRA2018,LatuneQST2019,ManzanoNJP2019,pusuluk2020}.

Here we propose a general framework based on collisional models~\cite{ScaraniPRL2002,ZimanPRA2002,Karevski2009,PalmaGiovannetti,CiccarelloPRA2013,LorenzoPRA2015,Landi2014b,LorenzoPRL2015,Barra2015,Strasberg2016,Manzano2016,Ciccarello2017,PezzuttoNJP2016,cusumano2018entropy,CampbellPRA2018,GrossQST2018,MohammadyPRA2018,PezzuttoQST2019,RodriguesPRL2019,ManatulyPRE2019,StrasbergPRL2019,SeahPRE2019,CakmakPRA2019,GuarnieriPLA2020,garcia2020ibm,Li_2020,cilluffo2020collisional} which allows us to analyse, in a consistent thermodynamic sense as recently proven in Ref.~\cite{DeChiaraNJP2018}, the effect of classical and quantum correlations between reservoirs in the functioning of quantum thermal machines.

\begin{figure}[t]
\begin{center}
\includegraphics[width=0.7\columnwidth]{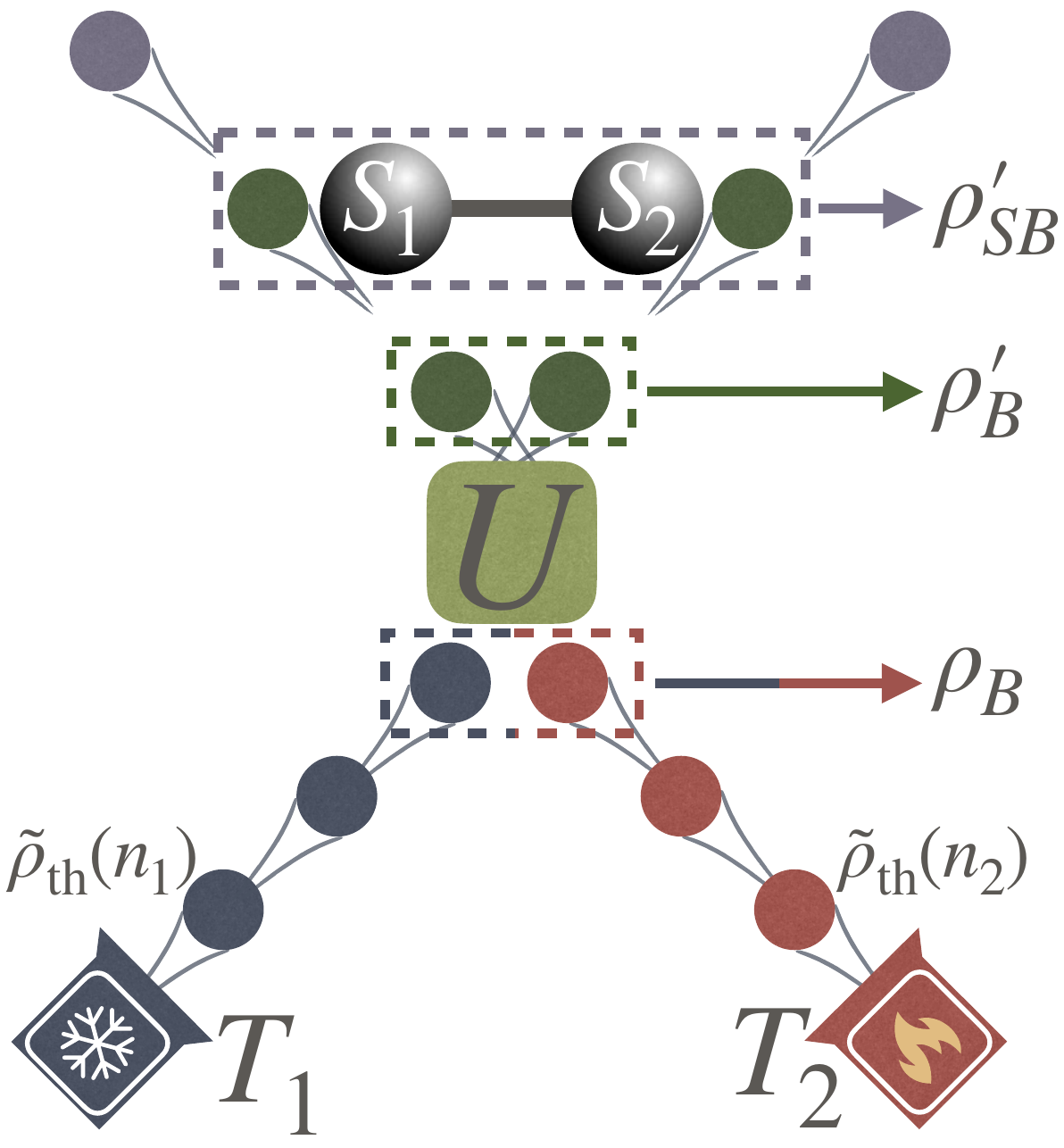}
\caption{Setup of our scheme: thermal particles emerging from the cold and hot baths and prepared in the states $\tilde\rho_{\rm th}(n_1)$ and $\tilde\rho_{\rm th}(n_1)$, respectively, are made to collide to each other under a unitary operation $U$. The emerging correlated particles in the global state $\rho_B'$ collide with the system's particles $S_1$ and $S_2$.}
\label{fig:setup}
\end{center}
\end{figure}

Our setup, depicted in Fig.~\ref{fig:setup}, consists of a working medium made of two quantum systems $S_1$ and $S_2$. Each of these is in contact with a reservoir modelled by the repeated interaction of flying auxiliary particles. These particles are first prepared in a thermal state at $T_1$ and $T_2$, respectively, and then undergo a unitary operation $U$ which correlates them before their collision with the systems $S_1$ and $S_2$. We study the steady state of the system after many collisions with the flying particles. 
 
Such a microscopic model has the advantage that all thermodynamic contributions, e.g. energy, heat, work and entropy, are accountable and that it is consistent with the laws of thermodynamics~\cite{Barra2015,Strasberg2016,DeChiaraNJP2018}. From the point of view of open quantum systems, our model leads to a non trivial evolution of a quantum system in contact with correlated reservoirs. As we will show, under the assumption of continuous evolution in the limit of collisions lasting an infinitesimal amount of time, the system's evolution can be cast in the form of a Markovian Lindblad master equation with collective jump operators acting non trivially on both systems (see also Ref.~\cite{DaryanooshPRA2018}).

In this paper we showcase the functioning of our model assuming the system and environment's particles to be qubits. However our framework is general and could be equally applied to higher-dimensional systems including infinite-dimensional ones, for instance quantum harmonic oscillators.

After introducing the setup in more detail in Sec.~\ref{sec:setup} and discussing its thermodynamics in Sec.~\ref{sec:thermodynamics}, we assume in Sec.~\ref{sec:swap} the unitary operation $U$ to be a partial swap. We will show how the amount of swapping between the flying qubits controls the amount and direction of the heat flow among the system's qubits. We then consider in Sec.~\ref{sec:random} the most general two-qubit unitary operations and study the distribution of work produced and heat  exchanged by the system when the unitaries are chosen randomly. We find that the extremal points of the distribution correspond to eight non-correlating unitaries with the optimal one corresponding to the complete two-qubit swap. In this context we analyse the quantum and classical correlations established among the quantum constituents of our setup. We find that, while correlations among the flying qubits are not necessary for the machine to work, quantum and classical correlations among the system's particle must be nonzero for the optimal performance.
Finally in Sec.~\ref{sec:conclusions} we summarise and conclude.

\section{Setup and preliminaries on  repeated interactions}
\label{sec:setup}
We assume the system to be composed of two coupled qubits described by the XXZ Hamiltonian:
\begin{equation}
H_S= J(\sigma_{x1}\sigma_{x2}+\sigma_{y1}\sigma_{y2}+\Delta \sigma_{z1}\sigma_{z2})+B_1\sigma_{z1}+B_2\sigma_{z2}
\end{equation}
where $J,\Delta$ and $B_i$ are the interaction strength, anisotropy and local magnetic field, respectively. Here and throughout the paper we assume $\hbar=k_B=J=1$ expressing all physical quantities in these units. The operators $\sigma_{xi},\sigma_{yi},\sigma_{zi}$ are the Pauli matrices for the qubit $i$. Notice that the total magnetisation $S_z=\sigma_{z1}+\sigma_{z2}$ is a conserved quantity as it commutes with the system Hamiltonian.

We assume the system to be affected by a reservoir modelled by the so-called repeated interactions~\cite{ScaraniPRL2002,ZimanPRA2002,Karevski2009,PalmaGiovannetti,CiccarelloPRA2013,LorenzoPRA2015,Landi2014b,LorenzoPRL2015,Barra2015,Strasberg2016,Manzano2016,Ciccarello2017,PezzuttoNJP2016,cusumano2018entropy,CampbellPRA2018,GrossQST2018,MohammadyPRA2018,PezzuttoQST2019,RodriguesPRL2019,ManatulyPRE2019,StrasbergPRL2019,SeahPRE2019,CakmakPRA2019,GuarnieriPLA2020,garcia2020ibm,Li_2020,cilluffo2020collisional}. In this model, represented in Fig.~\ref{fig:setup}, each qubit of the system interacts with a stream of uncoupled environment qubits (or flying qubits). The interaction between the system qubit and a flying qubit only lasts for a short time $\tau$ during which the interaction Hamiltonian is constant and given by:
\begin{equation}
H_{SB} =\sum_{i=1,2} \sqrt{\frac{\gamma(2n_i+1)}{2\tau}}(\sigma_{xi}\tilde\sigma_{xi}+\sigma_{yi}\tilde\sigma_{yi})
\end{equation}
where the operators $\tilde\sigma_{xi},\tilde\sigma_{yi},\tilde\sigma_{zi}$ are the Pauli matrices for a flying qubit interacting with the system qubit $i=1,2$.
The coefficient $\gamma$ determines the strength of the interaction while $n_i=(e^{2B_i\beta_i}-1)^{-1}$ models the thermal occupation of the flying qubit and is related to the inverse temperature $\beta_i=1/T_i$ of each bath. Additionally each flying qubit is subject to the local Hamiltonian:
\begin{equation}
H_{Bi}=B_i\tilde\sigma_{zi},\quad i=1,2,
\end{equation}
and we define $H_B=H_{B1}+H_{B2}$.

Usually, in the literature, the state of the environment qubits for different reservoirs has been assumed to be uncorrelated. In this paper we challenge this assumption and introduce some correlations, not necessarily quantum ones, between the flying qubits. These are initially uncorrelated and prepared in a thermal state corresponding to a thermal occupation $n_i$:
\begin{equation}
\label{eq:initialthermal}
\rho_B=\tilde\rho_{\rm th}(n_1)\otimes\tilde\rho_{\rm th}(n_2)
\end{equation}
 where
 \begin{equation}
\tilde \rho_{\rm th}(n_i)=\frac 12\left[\idop -(1+2n_i)^{-1}\tilde\sigma_{zi} \right]
\end{equation}
and $\idop$ is the identity operator.

We then correlate the flying qubits with a unitary transformation $U$ so that their state becomes:
\begin{equation}
\rho_B'=U\rho_BU^\dagger.
\end{equation}
After this initial preparation, each flying qubit collides with a system's qubit (see Fig.~\ref{fig:setup}). The collision, lasting for a time $\tau$, is described by the unitary operator:
\begin{equation}
U_{\rm collision}=e^{-i H_{\rm tot}\tau} 
\end{equation}
where $H_{\rm tot}=H_S+H_B+H_{SB}$ is the total Hamiltonian.

If we call $\rho_S(t)$ the state of the system at time $t$, then its state at time $t+\tau$ after the collision becomes
\begin{equation}
\rho_S(t+\tau)={\rm Tr}_B \left[U_{\rm collision}\,\rho_S(t)\otimes\rho_B'U_{\rm collision}^\dagger\right ].
\end{equation}
In the following we will always consider the steady state that the system approaches after many collisions which is defined by the relation:
\begin{equation}
\rho_S^{\rm steady}(t+\tau)=\rho_S^{\rm steady}(t)
\end{equation}
and for simplicity we will drop its time dependence and write simply $\rho_S^{\rm steady}$.

\section{Thermodynamics}
\label{sec:thermodynamics} 
We now discuss the different thermodynamics contributions arising in our setup after the system has reached its steady state. As the state of the system does not change anymore, the internal energy variation is zero:
\begin{equation}
\Delta E ={\rm Tr}\left[H_S\left(\rho_{SB}'-\rho_{SB}\right) \right]= 0
\end{equation}
where $\rho_{SB}=\rho_S^{\rm steady}\otimes\rho_B'$ and $\rho'_{SB}=U_{\rm collision}\;\rho_{SB}\,U^\dagger_{\rm collision}$.

To define heat and work we distinguish two scenarios, called {\it partial } and {\it complete} scenarios, which we explain in more details in the following.

\subsection{Partial scenario} 
In the partial scenario, we assume that we are provided with the flying qubits in the state $\rho_B'$ and that we are not paying for the work of the unitary $U$. In this scenario, we are provided with a non-equilibrium reservoir and thus the Clausius formulation of the second law of thermodynamics may not apply.
This is still an interesting scenario to study for two reasons: first, it continues the investigation of the functioning of thermodynamic machines  in the presence of non thermal environments~\cite{RossnagelPRL2014,LeggioPRA2015,LeggioPRE2016,Manzano2016,ReidEPL2017,AgarwallaPRB2017,AssisPRL2019,cherubim2019,PezzuttoQST2019,CarolloPRL2020}; second, it gives us an opportunity to study the open quantum system dynamics in the presence of correlated reservoirs as developed in Sec.~\ref{sec:swap}.

Under these assumptions the work is potentially produced or injected during the system-environment collision~\cite{DeChiaraNJP2018}. This is given by:
\begin{equation}
W_{\rm partial}={\rm Tr}\left[(H_S+H_B) \left(\rho_{SB}'-\rho_{SB}\right) \right]
\end{equation}
Similarly the heat exchanged by the system with the flying qubits is equal to the energy balance of the latter ones:
\begin{equation}
Q^{(i)}_{\rm partial}=-{\rm Tr}\left[H_{Bi}\left(\rho_{SB}'-\rho_{SB}\right) \right], \quad i=1,2.
\end{equation}

At steady state, the first law for the two-qubit system reads:
\begin{equation}
\Delta E = Q_{\rm partial}+W_{\rm partial}=0,
\end{equation}
where $Q_{\rm partial}=Q^{(1)}_{\rm partial}+Q^{(2)}_{\rm partial}$.
Notice that throughout this paper we employ the convention that positive work or heat corresponds to energy injected into the system contributing to the {\it increase} of the system energy.

An entropic formulation of the second law can  be derived in terms of the non-negativity of the entropy production (see for example Refs.~\cite{EspositoNJP2010,ReebNJP2014,PtaszynskiPRL2019} and Chapter 28 by R. Uzdin in Ref.~\cite{ThermoBook}):
\begin{equation}
\Sigma_{\rm partial} = I(\rho'_{SB})+ D(\rho''_B||\rho'_B) \ge 0.
\end{equation}
In the expression of $\Sigma$, we have used the mutual information between two quantum objects $O_1$ and $O_2$ defined as:
\begin{equation}
\label{eq:mi}
I_{O1O2}=S(\rho_{O1})+S(\rho_{O2})-S(\rho_{O1O2})
\end{equation}
where $S(\rho)=-{\rm Tr}\rho\ln\rho$ is the von Neumann entropy and $\rho_i, \, i=O_1,O_2,O_1O_2$ are the density matrices of the corresponding objects.
We have also used the relative entropy: 
$D(\rho||\sigma)={\rm Tr}\rho\ln\rho-{\rm Tr}\rho\ln\sigma$ 
and we have defined the state of the environment after the collision: $\rho''_B={\rm Tr}_S \rho'_{SB}$.
The non negativity of $\Sigma_{\rm partial}$ is ensured by the non negativity of the mutual information and the relative entropy.

\subsection{Complete scenario}
In the complete scenario we account for the extra work needed to implement the correlating unitary $U$:
\begin{equation}
W_U = {\rm Tr}\left[H_B\left(\rho_{B}'-\rho_{B}\right) \right]
\end{equation}
so that the total work in the complete scenario is the sum of the two contributions:
\begin{equation}
W_{\rm complete} = W_{\rm partial}+W_U.
\end{equation}

Equally, the heat exchanged is the energy balance of the environment during the whole process: from their preparation into the product thermal state $\rho_B$ to their final state after the two unitaries $U$ and  $U_{\rm collision}$:
\begin{equation}
Q^{(i)}_{\rm complete} = -{\rm Tr}\left[H_{Bi}\left(\rho_{SB}'-\rho^{\rm steady}_{S}\otimes \rho_B\right) \right], \quad i=1,2.
\end{equation}
and $Q_{\rm complete}=Q^{(1)}_{\rm complete}+Q^{(2)}_{\rm complete}$ which in general differs from $Q_{\rm partial}$.

A modified first law holds also in this scenario:
\begin{equation}
\Delta E = Q_{\rm complete}+W_{\rm complete}=0.
\end{equation}

It is possible to define the entropy production also in the complete scenario:
\begin{equation}
\Sigma_{\rm complete} = I(\rho'_{SB})+ D(\rho''_B||\rho_B) \ge 0.
\end{equation}
Moreover in this case, since $\rho_B$ is the tensor product of equilibrium states for each reservoir, see Eq.~\eqref{eq:initialthermal}, we can connect the change in entropy in the system $\Delta S$ with the entropy production and the heat flow:
\begin{equation}
\Delta S = \Sigma_{\rm complete}+\sum_{i=1}^2 \beta_i Q^{(i)}_{\rm complete}
\end{equation}
from which we can write Clausius inequality:
\begin{equation}
\Delta S -\sum_{i=1}^2 \beta_i Q^{(i)}_{\rm complete}= \Sigma_{\rm complete}\ge 0,
\end{equation}
where at steady state $\Delta S=0$.

\section{Reservoirs in a partially swapped locally thermal state}
\label{sec:swap}

Here, we consider the special case in which the unitary $U$ correlating the flying qubits is a partial swap operation:
\begin{equation}
\label{eq:USWAP}
S_\phi = \exp\left\{-i \frac \phi 2 (\tilde\sigma_{x1}\tilde\sigma_{y2}-\tilde\sigma_{y1}\tilde\sigma_{x2}) \right\}
\end{equation}
which is a total swap for $\phi=\pi/2$. This leads to the following expression for the bath density matrix after the action of the partial swap $S_\phi$:
\begin{widetext}
\begin{eqnarray}
\label{eq:UrhoB}
\rho'_B&=&\frac{1}{(1+2n_1)(1+2n_2)}\times
\\
&\times&
\left(\begin{array}{cccc}
n_1 n_2 & 0 & 0 & 0 \\
0 &\frac 12( n_1+n_2+2n_1n_2+(n_1-n_2)\cos2\phi)& (n_2-n_1)\sin\phi\cos\phi & 0 \\
0 & (n_2-n_1)\sin\phi\cos\phi &\frac 12( n_1+n_2+2n_1n_2-(n_1-n_2)\cos2\phi )& 0 \\
0 & 0 & 0 & (1+n_1)(1+n_2)
\end{array}\right)
\end{eqnarray}
\end{widetext}
where the basis of eigenstates of $\sigma_{z1}$ and $\sigma_{z2}$ has been used to write the matrix representation of $\rho'_B$.

\begin{figure}[t]
\begin{center}
\includegraphics[width=0.7\columnwidth]{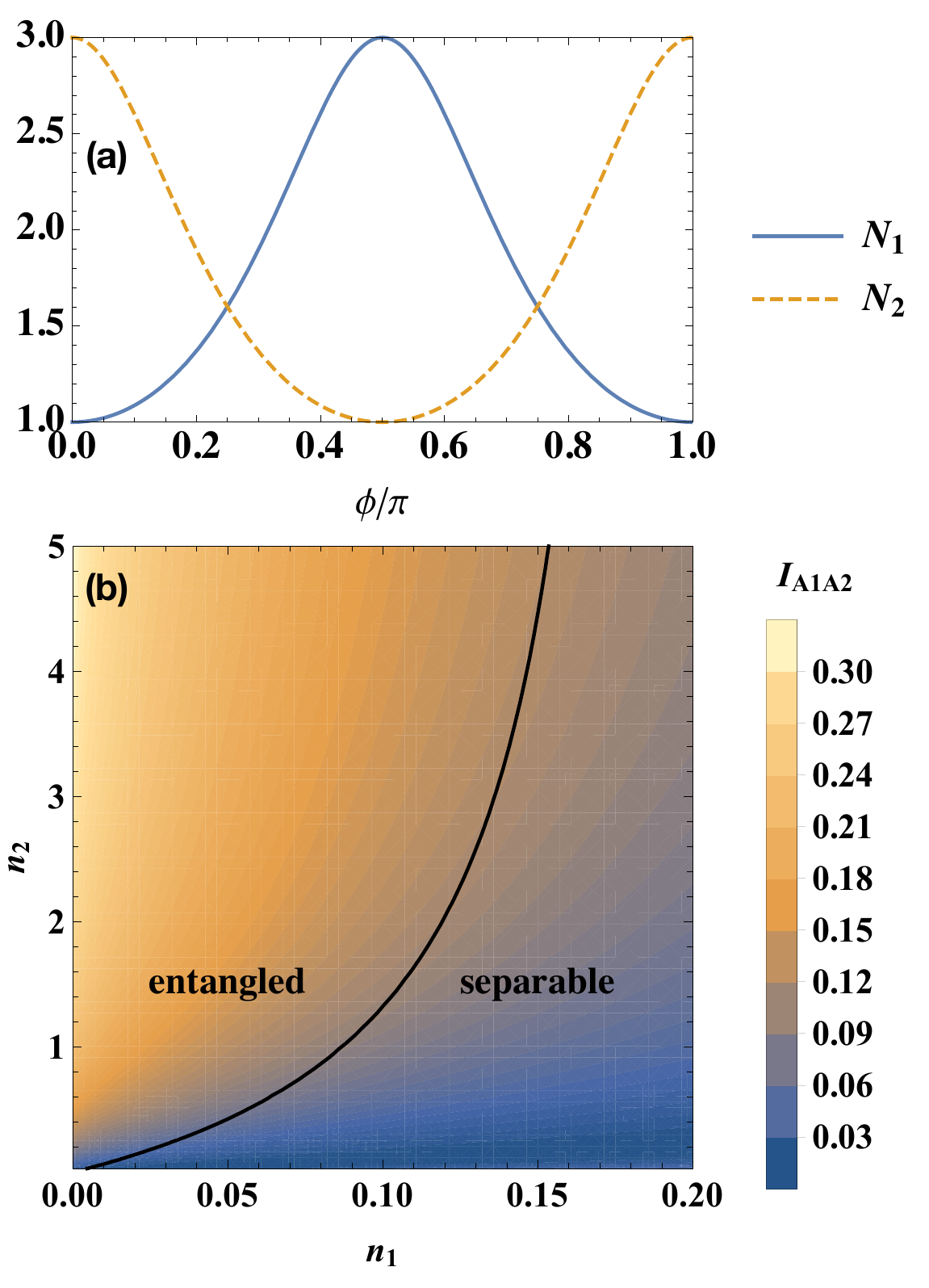}
\caption{{\bf (a)}: Effective populations of the two flying qubits as a function of $\phi$ for $n_1=1$ and $n_2=3$. {\bf (b)} Mutual information between the two flying qubits as a function of $n_1$ and $n_2$ for $\phi=0.2\pi$. The solid line corresponds to zero entanglement and divides separable and entangled states.}
\label{fig:effectivepopulationsmi}
\end{center}
\end{figure}

The action of the partial swap is to partially exchange populations between the auxiliary qubits. Indeed, the two flying qubits are still locally in a thermal state although at a different temperature compared to their state before the partial swap:
 \begin{equation}
\tilde\rho'_{i}={\rm Tr}_{\bar i}\; \rho'_B= \tilde\rho_{th}(N_i)
\end{equation}
where $i=1,2$ and ${\rm Tr}_{\bar i}$ represents the partial trace with respect to the qubit other than $i$.
The effective population after the application of $S_\phi$ is:
\begin{equation}
N_i=\frac 12 \frac{n_1+n_2+4n_1n_2+(-1)^i (n_2-n_1)\cos2\phi}
{1+n_1+n_2-(-1)^i (n_2-n_1)\cos2\phi}
\end{equation}
The effective populations $N_i$ of the two flying qubits are shown in Fig.~\ref{fig:effectivepopulationsmi} where it is evident that for $\phi=\pi/2$ the populations are completely swapped.

The partial swap creates classical and quantum correlations, measured for instance by the mutual information $I_{A1A2}$ of $\rho'_B$ among the two auxiliary qubits.
The mutual information  is always nonzero as shown in Fig.~\ref{fig:effectivepopulationsmi} except for the special values of $\phi=m\pi/2, \; m\in \mathbb{Z}$, regardless of the values $n_1$ and $n_2$. This happens because, as it is evident from Eq.~\eqref{eq:UrhoB}, $\rho'_B$ is not a product state.
Notice how, in Fig.~\ref{fig:effectivepopulationsmi}, unbalancing the two populations $n_1$ and $n_2$ leads to larger correlations. 

The work for implementing the partial swap is given by:
\begin{eqnarray}
\label{eq:WUswap}
W_U &=& -\frac{2(B_1-B_2)(n_1-n_2)\sin^2\phi}{(1+2n_1)(1+2n_2)}.
\end{eqnarray}
Similarly to what is found in Ref.~\cite{DeChiaraNJP2018, CampisiNJP2015}, the work vanishes for $B_1=B_2$ or $n_1=n_2$ and is maximum for the total swap $\phi=\pi/2$.

We set the parameters in such a way that for $\phi=0$ the two-qubit system acts as an engine ($W<0, Q^{(1)}<0,Q^{(2)}>0$ for both scenarios). It is possible to find the analytical expression of the steady state, the work and heats exchanged per cycle. While their expressions are quite long and hard to read, it is possible to show that all their ratios are proportional to ratios of the local magnetic fields, $B_1$ and $B_2$. As a consequence, and similarly to other models~\cite{CampisiNJP2015,DeChiaraNJP2018,HewgillPRA2018}, the efficiency $\eta=|W|/Q^{(2)}$, if the setup works as an engine, or the coefficient of performance (COP) $\eta_{\rm COP}=Q^{(1)}/W$, if the setup works as a refrigerator, correspond to the Otto values. For example in the former case, the efficiency is given by ($B_2<B_1$):
\begin{equation}
\label{eq:etaotto}
\eta=1-\frac{B_2}{B_1}
\end{equation}
and in the latter case the COP is given by ($B_1<B_2$):
\begin{equation}
\eta_{\rm COP}=\frac{B_1}{B_2-B_1}
\end{equation}

In the partial scenario, $\phi$ can be used to control the functioning of the thermodynamic machine making it switch from an engine to a refrigerator. In the interval $0\le \phi \le \pi/2$, the machine behaves as an engine for $0\le \phi < \pi/4$ and as a refrigerator for $\pi/4< \phi \le \pi/2$. For values of $\phi$ outside the interval $[0,\pi/2]$, the situation just described repeats periodically. The switching points are found for $N_1=N_2$ which occurs for $\cos 2\phi=0$, i.e. $\phi=(2m+1) \pi/4, m\in \mathbb{Z}$.
These switching points correspond to effective Carnot points where work and heat exchanged are zero (see also \cite{DeChiaraNJP2018} where the condition $n_1=n_2$ corresponds to the Carnot point).

\begin{figure}[t]
\begin{center}
\includegraphics[width=0.8\columnwidth]{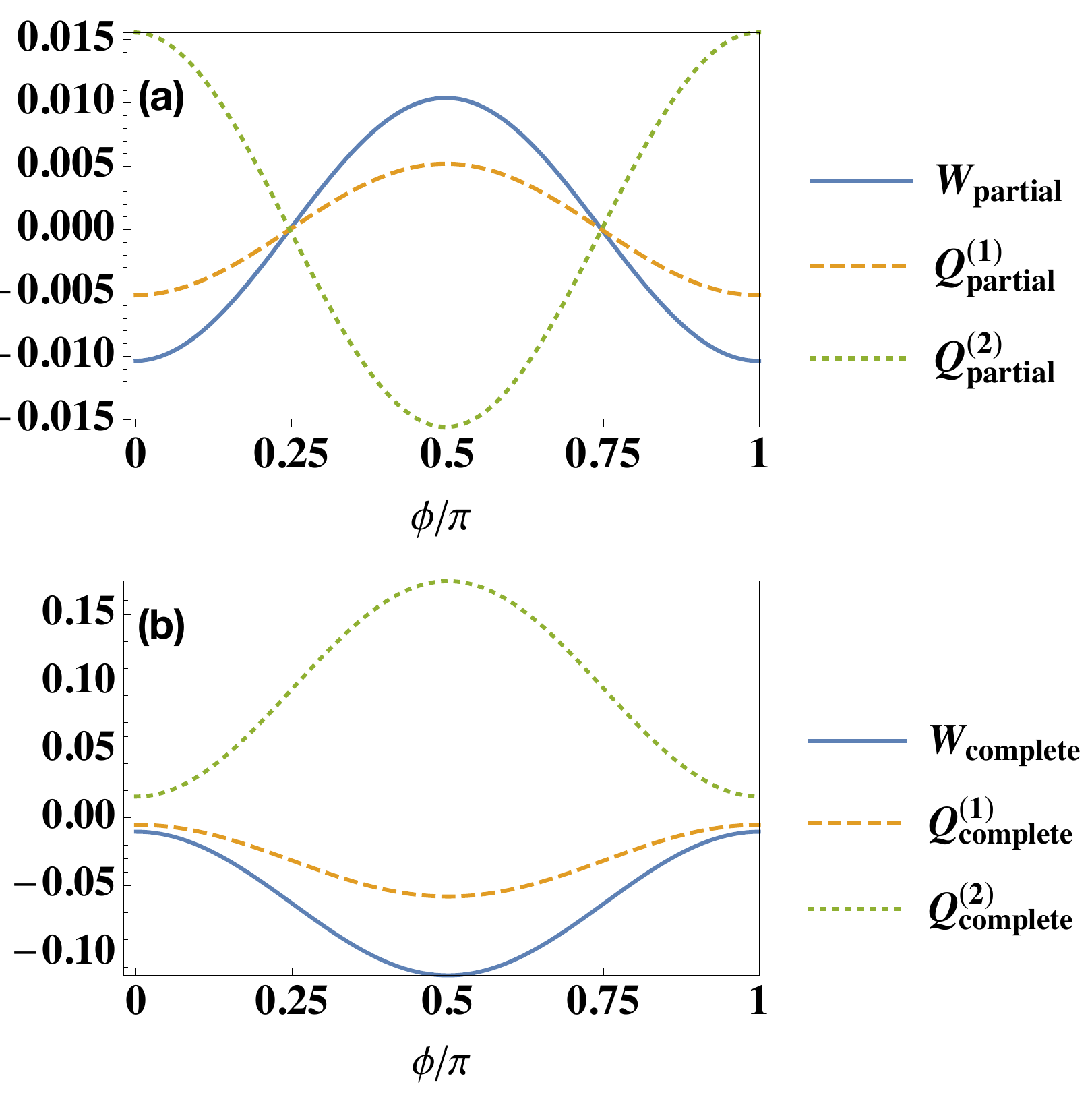}
\caption{Thermodynamic quantities: work and heat exchanged per collision for the steady state of the system in contact with the flying qubits prepared with the partial swap Eq.~\eqref{eq:USWAP}. Top: partial scenario. Bottom: complete scenario. Parameters: $\Delta=1$, $\gamma=1$, $B_1=0.1$, $B_2=0.3$, $n_1=0.1, n_2=2,\tau=0.1$.}
\label{fig:workSWAP}
\end{center}
\end{figure}

Numerical results for the partial scenario are plotted in Fig.~\ref{fig:workSWAP} which shows the periodic change of the machine operating as an engine and as a refrigerator. Interestingly, the largest absolute values of the thermodynamic quantities are obtained for the values $\phi=m \pi/2, \; m\in \mathbb{Z}$ at which the partial swap operation $S_\phi$ corresponds to the identity (even $m$) and the total swap (odd $m$).

Thus the partial swap operation can be employed as a valve to control the direction of the heat flowing between the system's qubits without altering the equilibrium reservoirs that prepare the flying qubits. These can be particularly useful in physical implementations in which one does not have full control of the environment.

Let us now move to the complete scenario. In this case, since we are taking into account all work contributions including those coming from the action of the swap operation, both first and second laws, the latter expressed in terms of heat exchanged, are fulfilled. The corresponding results plotted in Fig.~\ref{fig:workSWAP} show that the work and heats exchanged do not change sign and the machine always behaves as an engine. The action of the partial swap is to amplify the work production and the heats exchanged, while keeping the same performance, reaching a maximum for $\phi=m \pi/2, m=\pm 1,3,5, \dots$ at which the swap is total. 

This shows that given the two flying qubits, initially in equilibrium states with fixed temperatures, the maximum work that can be extracted is achieved by completely swapping their states before making them to collide with the system's qubit. As we will see in Sec.~\ref{sec:random} this is the maximum value that can be obtained for any unitary operation between the flying qubits.

We end up this section by considering the continuous limit of $\tau\to 0$. Up to first order in $\phi$ one obtains the following Lindblad master equation:
\begin{eqnarray}
\label{eq:me}
\dot\rho_S &=& -i[H_S,\rho_S]+\gamma\sum_{i=1,2} (1+n_i)\mathcal L_{\sigma_{-i}}(\rho_S)+n_i\mathcal L_{\sigma_{+i}}(\rho_S)
\\
&+&\frac{\gamma \phi(n_2-n_1)\left[\mathcal M(\sigma_{+1},\sigma_{-2},\rho_S)+\mathcal M(\sigma_{-2},\sigma_{+1},\rho_S)+h.c.\right]}{\sqrt{(1+2n_1)(1+2n_2)}}
\nonumber
\end{eqnarray}
where $\mathcal L_{a}(\rho) = 2a\rho a^\dagger-a^\dagger a \rho-\rho a^\dagger a$ is the usual Lindblad operator and $\mathcal M(a,b,\rho) = 2a\rho b-b a \rho-\rho b a$ is a modified Lindblad-like operator. In Eq.~\eqref{eq:me} we have also defined the jump operators $\sigma_{\pm i} =\frac 12 (\sigma_{xi}\pm i \sigma_{yi})$.

The presence of quantum correlations in the initial state of the bath is the reason for the appearance in the master equation of the collective term $\mathcal M$  proportional to $\phi$ which corresponds to environment-induced processes of emission of an excitation from one qubit and absorption from the other. 

In this continuous limit, it is possible to define work power and heat currents both in the partial and complete scenario. However, in the complete scenario, one should assume that the work necessary for the swap operation given in Eq.~\eqref{eq:WUswap} scales to zero as $\tau\to 0$. This is indeed the case up to first order in $\phi$ consistently with our master equation. 

\section{Reservoirs prepared by random unitaries}
\label{sec:random}
Here we generalise the approach developed in the previous section by considering thermal flying qubits which are subject, before the collision with the system qubits, to a unitary transformation $U_R$. The unitary transformation is always the same for all collisions for a given setup and allows the system to reach a steady state which we then analyse. We repeat the same procedure for an ensemble of $6\times 10^6$ random unitaries $U_R$ chosen with uniform distribution according to the Haar measure \cite{Ozols}.

For a generic $U_R$, the creation of finite quantum coherences and correlations between the environment qubit may lead to difficulties in the derivation of a consistent master equation in the continuous limit $\tau\to 0$, see Ref.~\cite{RodriguesPRL2019}. 
In this section, we will therefore restrict our analysis to a small but finite $\tau$. Under this assumption, the repeated interactions remain a discrete map for the system which, after many applications reaches a steady state, which we analyse. 

\begin{figure}[t]
\begin{center}
\includegraphics[width=0.8\columnwidth]{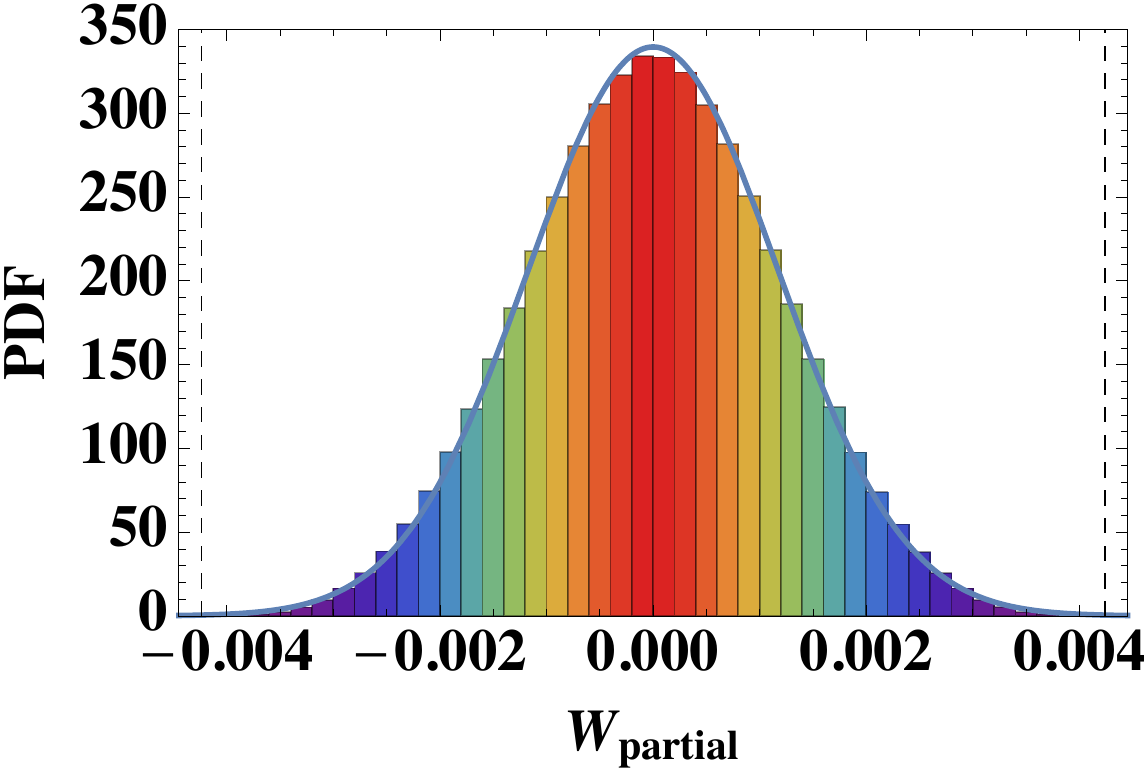}
\caption{Histogram of the partial work $W_{\rm partial}$ with the vertical axis measuring the corresponding probability density function (PDF) for $6\times 10^6$ random unitaries. The solid line is the best fit normal distribution with average $9.6 \times 10^{-7}$ and standard deviation $1.1\times 10^{-3}$. The vertical dashed lines signpost the minimum and maximum possible values of  $W_{\rm partial}$ obtained with non-correlating unitaries (see text). Parameters as in Fig.~\ref{fig:workSWAP} except for the local magnetic fields which are $B_1=0.1$ and $B_2=0.15$.}
\label{fig:randomworkpartial}
\end{center}
\end{figure}

We consider a setup which, for $U_R= \mathds 1$, corresponding to the situation in which no unitary is applied to the flying qubits before their collision with the system, operates as an engine. Analogous results can be obtained for the refrigerator regime. 

In the partial case, as in Sec.~\ref{sec:swap}, work and heat fluxes are proportional to each other with the prefactor being a ratio of the local magnetic fields, $B_1$ and $B_2$,  related to the Otto efficiency. For this reason we only show the analysis of the work probability distribution shown in Fig.~\ref{fig:randomworkpartial}. Its probability density function (PDF) is approximately Gaussian with a very small average statistically compatible with zero. 

The maximum and minimum values of the partial work $W_{\rm partial}$ are obtained with unitaries that do not create correlations, classical or quantum, among the auxiliary qubits. As we will see in more detail in the complete scenario, this type of unitary operations do not create any mutual information between the flying qubits but simply rearrange the populations of the four basis states. In the partial scenario we find that the minimum negative value of the work (largest value of produced work) is obtained with the unitary that inverts the populations of the qubit prepared in the hottest temperature and leaves unchanged the qubit prepared in the coldest temperature: $U_R=\sigma_{x2}$ (corresponding to operation III defined later). In contrast, the maximum positive value of the work (largest value of the work injected), is obtained with the unitary that inverts the populations of the qubit prepared in the coldest temperature and leaves unchanged the qubit prepared in the hottest temperature: $U_R=\sigma_{x1}$ (corresponding to operation VI defined later).

\begin{figure}[t]
\begin{center}
\includegraphics[width=0.99\columnwidth]{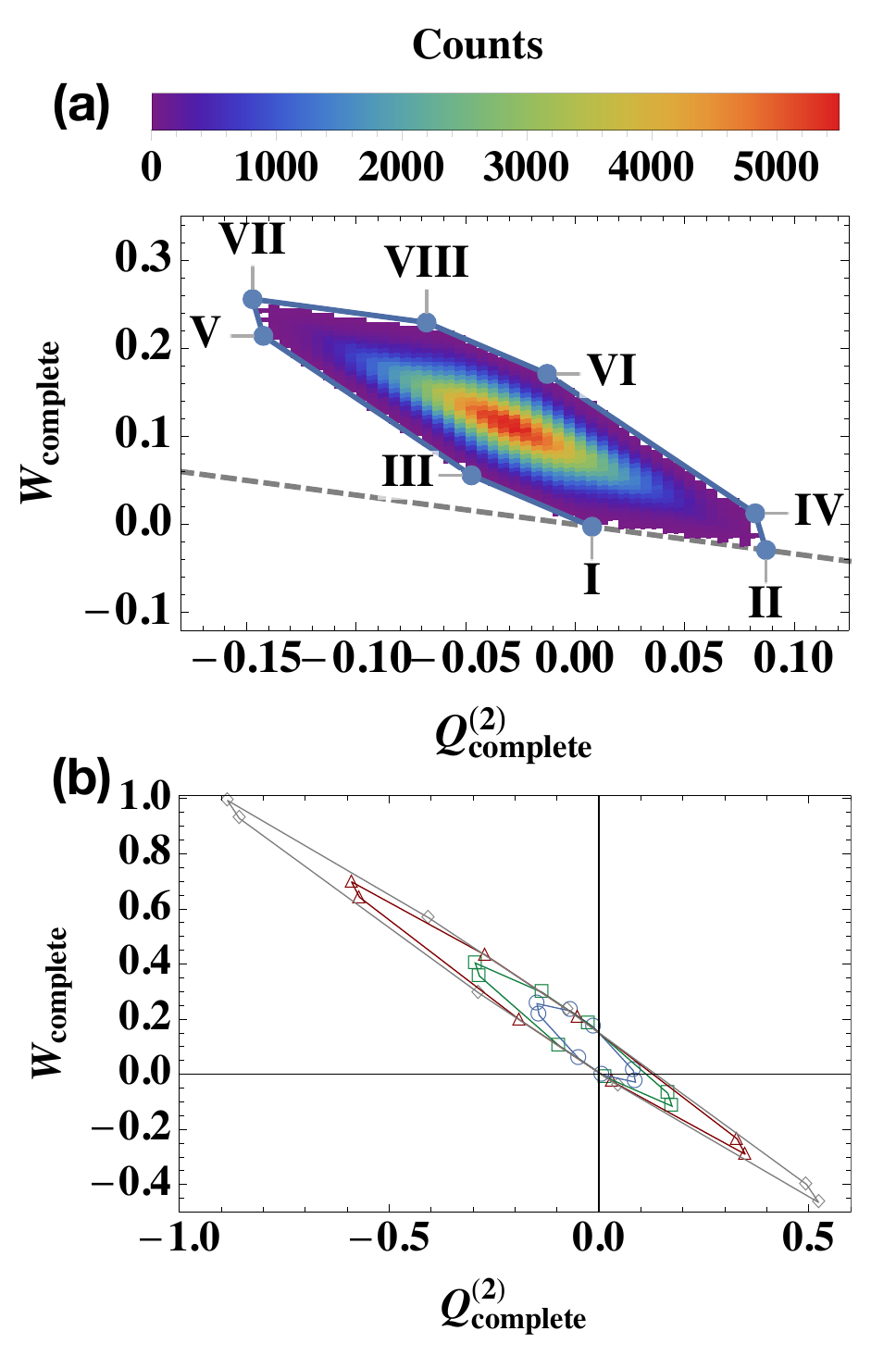}
\caption{{\bf (a)}: Joint histogram of the complete work $W_{\rm complete}$ versus heat $Q_{\rm complete}^{(2)}$. The dashed gray line represents the Otto efficiency. The blue dots, connected by thin solid lines, are obtained for non correlating unitaries marked by the corresponding number. These special points form an extremal octagon within which all other machines obtained for random unitaries must be located. The colour code identifies the number of values in each bin of the histogram. Parameters: $\Delta=1$, $\gamma=1$, $B_1=0.1$, $B_2=0.15$, $n_1=0.1, n_2=2,\tau=0.1$. {\bf (b)}: Extremal octagons obtained for the same parameters of the top panel but different values of $B_2=0.15,0.3,0.6,0.9$, corresponding to vertexes denoted by  circles, squares, triangles and diamonds, respectively.}
\label{fig:randomheatwork}
\end{center}
\end{figure}

Let us now pass to the complete scenario in which the work produced or extracted is not necessarily proportional to the two heat fluxes. The results show that, by applying different unitary operations, the different modes (accelerator, heater or heat engine) can be achieved with different efficiencies. We plot in Fig.~\ref{fig:randomheatwork} the joint histogram of the complete work and the heat input from the hot environment.  
The distribution is confined by a non-regular octagon whose eight vertexes correspond to unitary operations that do not create quantum or classical correlations between the environment qubits. Let $\{p_1,p_2,p_3,p_4\}$ be the vector of the populations of the density matrix $\rho_B$ of the flying qubits after their preparation in equilibrium states but before undergoing the unitary $U$. These eight unitary operations only affect the populations of the flying qubits according to:\\
\begin{center}
\begin{tabular}{ccc}
\hline
Label & Populations & Unitary
\\\hline\hline
{\rm I}& $\{p_1,p_2,p_3,p_4\}$& $\idop_1\idop_2$
\\\hline
{\rm II} & $\{p_1,p_3,p_2,p_4\}$& $ S_{\pi/2}$
\\\hline
{\rm III} & $\{p_2,p_1,p_4,p_3\}$& $ \idop_1 \sigma_{x2}$
\\\hline
{\rm IV} &$ \{p_2,p_4,p_1,p_3\}$&  $S_{\pi/2}\idop_1 \sigma_{x2}$
\\\hline
{\rm V} & $\{p_3,p_1,p_4,p_2\}$&   $S_{\pi/2} \sigma_{x1}\idop_2$
\\\hline
{\rm VI} &$ \{p_3,p_4,p_1,p_2\}$& $ \sigma_{x1}\idop_2$
\\\hline
{\rm VII} & $\{p_4,p_2,p_3,p_1\}$&  $S_{\pi/2}\sigma_{x1}\sigma_{x2}$
\\\hline
{\rm VIII} &$ \{p_4,p_3,p_2,p_1\}$& $ \sigma_{x1}\sigma_{x2}$
\\\hline
\end{tabular}
\end{center}
where $S_{\pi/2}$ is the transformation that completely swap the state of the two qubits, see Eq.~\eqref{eq:USWAP} for $\phi=\pi/2$.

Notice the symmetry between the unitary operations and the position of the corresponding vertex of the octagon. In Fig.~\ref{fig:randomheatwork}, two vertexes are opposite on the octagon if their transformations can be obtained one from the other by applying the total swap operation.
Moreover, one of the sides of the octagon, delimited by the vertexes I and II, corresponds to the Otto efficiency given in Eq.~\eqref{eq:etaotto}, so that points above the line corresponds to engines operating at a lower efficiency or, for $W_{\rm complete}\ge 0$ operating as an accelerator, if  $Q^{(2)}_{\rm complete}>0$ or a heater if $Q^{(2)}_{\rm complete}<0$. In the complete scenario, there are no unitaries that lead to refrigeration, similarly to what  we discussed in Sec.~\ref{sec:swap} for the partial swap. 

The optimal point II yielding the larger amount of work produced (minimum negative value) corresponds to the complete swap operation. This can be related to an engine proposed by Campisi, Pekola and Fazio also based on a complete swap transformation but  applied to the qubits of the working fluid~\cite{CampisiNJP2015}. 

In Fig.~\ref{fig:randomheatwork} we also show how the shape and size of the extremal octagon change when $B_2$ is varied while all other parameters are kept  fixed. Notice that, as we increase $B_2$, the region corresponding to the engine increases, leading to a higher probability of achieving this operation mode. This can be understood as follows. First, $Q^{(2)}_{\rm complete}\propto B_2$, thus if we rescaled the horizontal axis of the lower diagram in Fig.~\ref{fig:randomheatwork},  all the vertexes would have the same horizontal  coordinates. Second, for $W_{\rm complete}=W_{\rm partial}+W_U$ the situation is more involved. While the partial work always fulfils $W_{\rm partial}\propto B_2-B_1$,  the work $W_U$ needed to implement the non-correlating unitaries is a linear function of $B_2$ and $B_1$ which depends on the actual transformation. 
Therefore there exists no rescaling of the vertical axis that would bring the vertexes of different octagons to the same vertical coordinate. 
Furthermore, the vertex IV, corresponding to an accelerator for $B_2=0.15$, turns into an engine for the larger values of $B_2$ we analysed. 

To get more insight into the relationship between the functioning of the whole system as an engine and the quantum features of the working medium quantum steady state, we looked at two measures of correlation between two parts of the engine.
The first is the mutual information defined in Eq.~\eqref{eq:mi}.
The mutual information quantifies both classical and quantum correlations. If one subtracts the maximum amount of classical correlations that can be obtained by local measurements, the quantum discord, a genuine measure of quantum correlations, is obtained  \cite{PhysRevLett.88.017901, henderson2001classical} . This can be defined as:
 \begin{equation}
D_{O1O2}=I_{O1O2}-J_{O1O2},
 \end{equation}
where the classical information $J_{O1O}$ is the maximum information that can be extracted on $O_2$ if we perform local measurements on $O_1$:
 \begin{equation}
 \label{eq:JO1O2}
J_{O1O2}=S(\rho_{O2})-\min_{\{\Pi_i\}}\sum_{i=1}^{N}q_i S(\tilde\rho_i).
\end{equation}
In Eq.~\eqref{eq:JO1O2}, we have defined the probabilities $q_i ={\rm Tr}[\Pi_i\rho_{O1O2}\Pi_i]$ of the outcome $i$ and the post-measurement states $\tilde\rho_i= {\rm Tr}_{O1}[\Pi_i\rho_{O1O2}\Pi_i]$ of the object $O_2$. The minimisation is done over all possible sets of measurements $\{\Pi_i\}$ on $O_1$, not necessarily orthogonal projectors.

\begin{figure*}[htbp]
\begin{center}
\includegraphics[width=1.99\columnwidth]{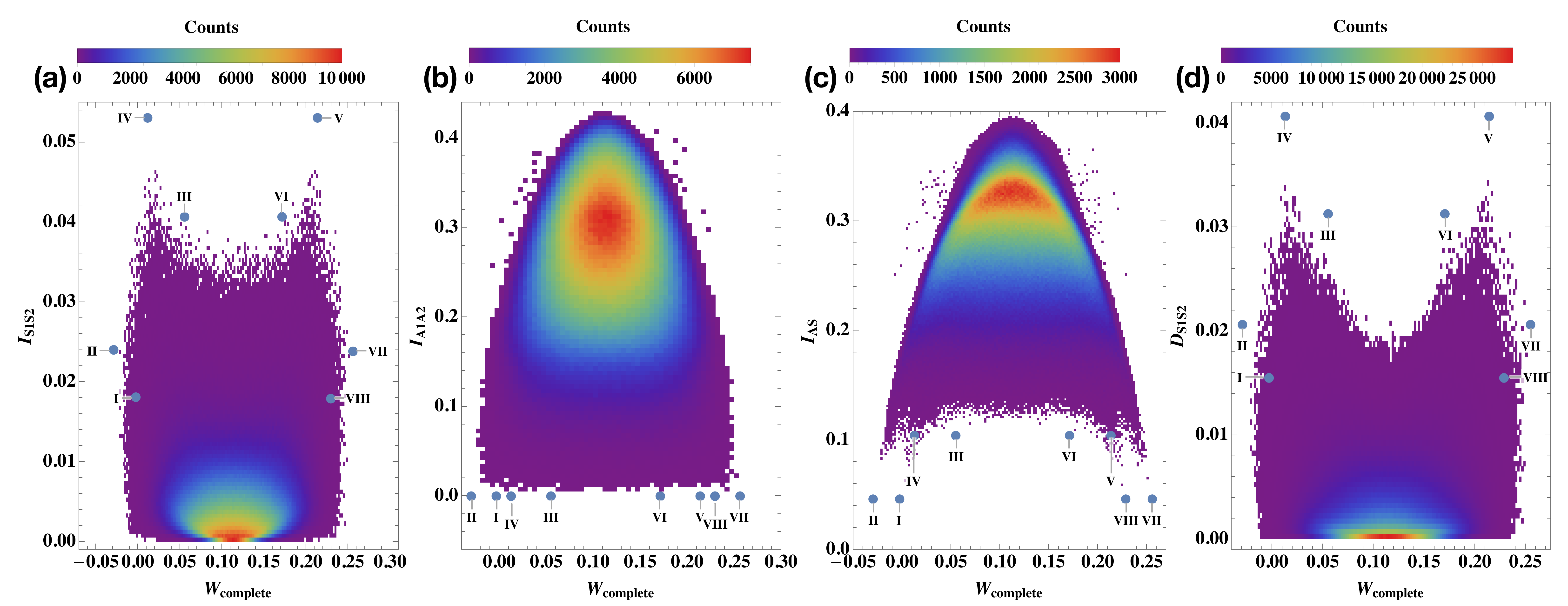}
\caption{Histograms of the mutual information and quantum discord versus the total work $W_{\rm complete}$. From left to right: {\bf (a)} $I_{S1S2}$, {\bf (b)} $I_{A1A2}$, {\bf (c)} $I_{AS}$, {\bf (d)} $D_{S1S2}$ (see definition in the main text). The blue dots are obtained for non correlating unitaries specified by their label I-VIII.}
\label{fig:randomquantumwork}
\end{center}
\end{figure*}

 The results are shown in Fig.~\ref{fig:randomquantumwork}. We start with the distribution of the mutual information $I_{S_1S_2}$ between the system qubits in the steady state. This shows that, although the optimal point II corresponds to uncorrelated flying qubits, the system qubits are nevertheless correlated as a result of their direct interaction and of reaching a non-equilibrium steady state due to the multiple collisions with the environment. A similar distribution is also obtained for the quantum discord $D_{S_1S_2}$, which shows that the system qubits are also genuinely quantum correlated. 
 
 We have also considered the mutual information $I_{A_1A_2}$ between the flying qubits after the unitary but before the collision with the system. This shows that the preparation unitaries applied before the correlations do indeed create a lot of correlations, quantum or classical, but these do not necessarily lead to large values of work produced or injected. The extremal operation II in fact corresponds to zero mutual information between the flying qubits, which are thus in a product state. Similar  conclusions, although quantitatively different, are reached when analysing concurrence and discord between the two flying qubits.
 
 Finally, we have analysed the mutual information $I_{AS}$ between the flying qubits and the system after the collision. As before, the state of the system is steady. Thus, although it does not change during the collision, it sustains correlations to be created between system and each pair of environmental qubits. These correlations are necessary for allowing exchange of heat  between the two reservoirs through the system. The distribution of $I_{AS}$ plotted in Fig.~\ref{fig:randomquantumwork}, shows that to achieve large amounts of work, and consequently heat exchange, it is sufficient a small value of  $I_{AS}$. In particular the non-correlating operations I-VIII, that leave the flying qubits in a product state, correspond to the smallest values of  $I_{AS}$.
 Finally, in all these figures of merit, the non-correlating operations I-VIII do not necessarily correspond to extremal points since these functions, mutual information and discord, are not linear functions of the state in contrast to the average work and heat.

\section{Conclusions}
\label{sec:conclusions}
In this paper we have proposed a general framework to model quantum thermal machines in contact with correlated reservoirs using repeated interactions. Different conclusions are found depending on whether we assume the partial or the complete scenario, the latter one being always consistent with the laws of thermodynamics. 

We have shown how, in the partial scenario, the amount of partial swapping among the flying qubits, allows one to control the operating mode of the thermal machine switching it from an engine to a refrigerator. 

In the case of random unitaries, we found a complex geometrical structure in the distribution of heat and work bounded by a non regular octagon whose vertexes correspond to non correlating unitaries.  We analysed the discord and mutual information between the flying qubits, the system qubits and between system and environment. We found that correlations in the system steady state and between system and environment are necessary to achieve the optimal performance.

Our work leads the way to future studies of open quantum systems with correlated environments consistently with thermodynamics.

\acknowledgements
We thank Marco Cattaneo, Francesco Ciccarello, Adam Hewgill and Gabriel Landi for useful discussions. 
GDC thanks the CNRS and the group Theory of Light-Matter and Quantum Phenomena of the Laboratoire Charles Coulomb for hospitality during his stay in Montpellier. We acknowledge support from the UK EPSRC EP/S02994X/1.
All the numerical data presented in this work can be found in Ref.~\cite{datarep}.

\bibliography{Refs}

\begin{thebibliography}{68}%
\makeatletter
\providecommand \@ifxundefined [1]{%
 \@ifx{#1\undefined}
}%
\providecommand \@ifnum [1]{%
 \ifnum #1\expandafter \@firstoftwo
 \else \expandafter \@secondoftwo
 \fi
}%
\providecommand \@ifx [1]{%
 \ifx #1\expandafter \@firstoftwo
 \else \expandafter \@secondoftwo
 \fi
}%
\providecommand \natexlab [1]{#1}%
\providecommand \enquote  [1]{``#1''}%
\providecommand \bibnamefont  [1]{#1}%
\providecommand \bibfnamefont [1]{#1}%
\providecommand \citenamefont [1]{#1}%
\providecommand \href@noop [0]{\@secondoftwo}%
\providecommand \href [0]{\begingroup \@sanitize@url \@href}%
\providecommand \@href[1]{\@@startlink{#1}\@@href}%
\providecommand \@@href[1]{\endgroup#1\@@endlink}%
\providecommand \@sanitize@url [0]{\catcode `\\12\catcode `\$12\catcode
  `\&12\catcode `\#12\catcode `\^12\catcode `\_12\catcode `\%12\relax}%
\providecommand \@@startlink[1]{}%
\providecommand \@@endlink[0]{}%
\providecommand \url  [0]{\begingroup\@sanitize@url \@url }%
\providecommand \@url [1]{\endgroup\@href {#1}{\urlprefix }}%
\providecommand \urlprefix  [0]{URL }%
\providecommand \Eprint [0]{\href }%
\providecommand \doibase [0]{http://dx.doi.org/}%
\providecommand \selectlanguage [0]{\@gobble}%
\providecommand \bibinfo  [0]{\@secondoftwo}%
\providecommand \bibfield  [0]{\@secondoftwo}%
\providecommand \translation [1]{[#1]}%
\providecommand \BibitemOpen [0]{}%
\providecommand \bibitemStop [0]{}%
\providecommand \bibitemNoStop [0]{.\EOS\space}%
\providecommand \EOS [0]{\spacefactor3000\relax}%
\providecommand \BibitemShut  [1]{\csname bibitem#1\endcsname}%
\let\auto@bib@innerbib\@empty
\bibitem [{\citenamefont {Binder}\ \emph {et~al.}(2018)\citenamefont {Binder},
  \citenamefont {Correa}, \citenamefont {Gogolin}, \citenamefont {Anders},\
  and\ \citenamefont {Adesso}}]{ThermoBook}%
  \BibitemOpen
  \bibinfo {editor} {\bibfnamefont {F.}~\bibnamefont {Binder}}, \bibinfo
  {editor} {\bibfnamefont {L.~A.}\ \bibnamefont {Correa}}, \bibinfo {editor}
  {\bibfnamefont {C.}~\bibnamefont {Gogolin}}, \bibinfo {editor} {\bibfnamefont
  {J.}~\bibnamefont {Anders}}, \ and\ \bibinfo {editor} {\bibfnamefont
  {G.}~\bibnamefont {Adesso}},\ eds.,\ \href {\doibase
  https://doi.org/10.1007/978-3-319-99046-0} {\emph {\bibinfo {title}
  {Thermodynamics in the Quantum Regime}}}\ (\bibinfo  {publisher} {Springer},\
  \bibinfo {year} {2018})\BibitemShut {NoStop}%
\bibitem [{\citenamefont {Millen}\ and\ \citenamefont
  {Xuereb}(2016)}]{XuerebReview}%
  \BibitemOpen
  \bibfield  {author} {\bibinfo {author} {\bibfnamefont {J.}~\bibnamefont
  {Millen}}\ and\ \bibinfo {author} {\bibfnamefont {A.}~\bibnamefont
  {Xuereb}},\ }\href {http://stacks.iop.org/1367-2630/18/i=1/a=011002}
  {\bibfield  {journal} {\bibinfo  {journal} {New Journal of Physics}\ }\textbf
  {\bibinfo {volume} {18}},\ \bibinfo {pages} {011002} (\bibinfo {year}
  {2016})}\BibitemShut {NoStop}%
\bibitem [{\citenamefont {Goold}\ \emph {et~al.}(2016)\citenamefont {Goold},
  \citenamefont {Huber}, \citenamefont {Riera}, \citenamefont {del Rio},\ and\
  \citenamefont {Skrzypczyk}}]{GooldReview}%
  \BibitemOpen
  \bibfield  {author} {\bibinfo {author} {\bibfnamefont {J.}~\bibnamefont
  {Goold}}, \bibinfo {author} {\bibfnamefont {M.}~\bibnamefont {Huber}},
  \bibinfo {author} {\bibfnamefont {A.}~\bibnamefont {Riera}}, \bibinfo
  {author} {\bibfnamefont {L.}~\bibnamefont {del Rio}}, \ and\ \bibinfo
  {author} {\bibfnamefont {P.}~\bibnamefont {Skrzypczyk}},\ }\href
  {http://stacks.iop.org/1751-8121/49/i=14/a=143001} {\bibfield  {journal}
  {\bibinfo  {journal} {J. Phys. A: Math. Theor.}\ }\textbf {\bibinfo {volume}
  {49}},\ \bibinfo {pages} {143001} (\bibinfo {year} {2016})}\BibitemShut
  {NoStop}%
\bibitem [{\citenamefont {Vinjanampathy}\ and\ \citenamefont
  {Anders}(2016)}]{AndersReview}%
  \BibitemOpen
  \bibfield  {author} {\bibinfo {author} {\bibfnamefont {S.}~\bibnamefont
  {Vinjanampathy}}\ and\ \bibinfo {author} {\bibfnamefont {J.}~\bibnamefont
  {Anders}},\ }\href {\doibase 10.1080/00107514.2016.1201896} {\bibfield
  {journal} {\bibinfo  {journal} {Contemporary Physics}\ }\textbf {\bibinfo
  {volume} {57}},\ \bibinfo {pages} {545} (\bibinfo {year} {2016})}\BibitemShut
  {NoStop}%
\bibitem [{\citenamefont {Mitchison}(2019)}]{Mitchison2019}%
  \BibitemOpen
  \bibfield  {author} {\bibinfo {author} {\bibfnamefont {M.~T.}\ \bibnamefont
  {Mitchison}},\ }\href {\doibase 10.1080/00107514.2019.1631555} {\bibfield
  {journal} {\bibinfo  {journal} {Contemporary Physics}\ }\textbf {\bibinfo
  {volume} {60}},\ \bibinfo {pages} {164} (\bibinfo {year} {2019})}\BibitemShut
  {NoStop}%
\bibitem [{\citenamefont {Deffner}\ and\ \citenamefont
  {Campbell}(2019)}]{deffner2019quantum}%
  \BibitemOpen
  \bibfield  {author} {\bibinfo {author} {\bibfnamefont {S.}~\bibnamefont
  {Deffner}}\ and\ \bibinfo {author} {\bibfnamefont {S.}~\bibnamefont
  {Campbell}},\ }\href@noop {} {\emph {\bibinfo {title} {Quantum
  Thermodynamics: An introduction to the thermodynamics of quantum
  information}}}\ (\bibinfo  {publisher} {Morgan \& Claypool Publishers},\
  \bibinfo {year} {2019})\BibitemShut {NoStop}%
\bibitem [{\citenamefont {Ro{\ss}nagel}\ \emph {et~al.}(2016)\citenamefont
  {Ro{\ss}nagel}, \citenamefont {Dawkins}, \citenamefont {Tolazzi},
  \citenamefont {Abah}, \citenamefont {Lutz}, \citenamefont {Schmidt-Kaler},\
  and\ \citenamefont {Singer}}]{RossnagelScience2016}%
  \BibitemOpen
  \bibfield  {author} {\bibinfo {author} {\bibfnamefont {J.}~\bibnamefont
  {Ro{\ss}nagel}}, \bibinfo {author} {\bibfnamefont {S.~T.}\ \bibnamefont
  {Dawkins}}, \bibinfo {author} {\bibfnamefont {K.~N.}\ \bibnamefont
  {Tolazzi}}, \bibinfo {author} {\bibfnamefont {O.}~\bibnamefont {Abah}},
  \bibinfo {author} {\bibfnamefont {E.}~\bibnamefont {Lutz}}, \bibinfo {author}
  {\bibfnamefont {F.}~\bibnamefont {Schmidt-Kaler}}, \ and\ \bibinfo {author}
  {\bibfnamefont {K.}~\bibnamefont {Singer}},\ }\href {\doibase
  10.1126/science.aad6320} {\bibfield  {journal} {\bibinfo  {journal}
  {Science}\ }\textbf {\bibinfo {volume} {352}},\ \bibinfo {pages} {325}
  (\bibinfo {year} {2016})}\BibitemShut {NoStop}%
\bibitem [{\citenamefont {Ronzani}\ \emph {et~al.}(2018)\citenamefont
  {Ronzani}, \citenamefont {Karimi}, \citenamefont {Senior}, \citenamefont
  {Chang}, \citenamefont {Peltonen}, \citenamefont {Chen},\ and\ \citenamefont
  {Pekola}}]{ronzani2018tunable}%
  \BibitemOpen
  \bibfield  {author} {\bibinfo {author} {\bibfnamefont {A.}~\bibnamefont
  {Ronzani}}, \bibinfo {author} {\bibfnamefont {B.}~\bibnamefont {Karimi}},
  \bibinfo {author} {\bibfnamefont {J.}~\bibnamefont {Senior}}, \bibinfo
  {author} {\bibfnamefont {Y.-C.}\ \bibnamefont {Chang}}, \bibinfo {author}
  {\bibfnamefont {J.~T.}\ \bibnamefont {Peltonen}}, \bibinfo {author}
  {\bibfnamefont {C.}~\bibnamefont {Chen}}, \ and\ \bibinfo {author}
  {\bibfnamefont {J.~P.}\ \bibnamefont {Pekola}},\ }\href@noop {} {\bibfield
  {journal} {\bibinfo  {journal} {Nature Physics}\ }\textbf {\bibinfo {volume}
  {14}},\ \bibinfo {pages} {991} (\bibinfo {year} {2018})}\BibitemShut
  {NoStop}%
\bibitem [{\citenamefont {Maslennikov}\ \emph {et~al.}(2019)\citenamefont
  {Maslennikov}, \citenamefont {Ding}, \citenamefont {Habl{\"u}tzel},
  \citenamefont {Gan}, \citenamefont {Roulet}, \citenamefont {Nimmrichter},
  \citenamefont {Dai}, \citenamefont {Scarani},\ and\ \citenamefont
  {Matsukevich}}]{Maslennikov19}%
  \BibitemOpen
  \bibfield  {author} {\bibinfo {author} {\bibfnamefont {G.}~\bibnamefont
  {Maslennikov}}, \bibinfo {author} {\bibfnamefont {S.}~\bibnamefont {Ding}},
  \bibinfo {author} {\bibfnamefont {R.}~\bibnamefont {Habl{\"u}tzel}}, \bibinfo
  {author} {\bibfnamefont {J.}~\bibnamefont {Gan}}, \bibinfo {author}
  {\bibfnamefont {A.}~\bibnamefont {Roulet}}, \bibinfo {author} {\bibfnamefont
  {S.}~\bibnamefont {Nimmrichter}}, \bibinfo {author} {\bibfnamefont
  {J.}~\bibnamefont {Dai}}, \bibinfo {author} {\bibfnamefont {V.}~\bibnamefont
  {Scarani}}, \ and\ \bibinfo {author} {\bibfnamefont {D.}~\bibnamefont
  {Matsukevich}},\ }\href {\doibase 10.1038/s41467-018-08090-0} {\bibfield
  {journal} {\bibinfo  {journal} {Nature Communications}\ }\textbf {\bibinfo
  {volume} {10}},\ \bibinfo {pages} {202} (\bibinfo {year} {2019})}\BibitemShut
  {NoStop}%
\bibitem [{\citenamefont {von Lindenfels}\ \emph {et~al.}(2019)\citenamefont
  {von Lindenfels}, \citenamefont {Gr\"ab}, \citenamefont {Schmiegelow},
  \citenamefont {Kaushal}, \citenamefont {Schulz}, \citenamefont {Mitchison},
  \citenamefont {Goold}, \citenamefont {Schmidt-Kaler},\ and\ \citenamefont
  {Poschinger}}]{LindenfelsPRL2019}%
  \BibitemOpen
  \bibfield  {author} {\bibinfo {author} {\bibfnamefont {D.}~\bibnamefont {von
  Lindenfels}}, \bibinfo {author} {\bibfnamefont {O.}~\bibnamefont {Gr\"ab}},
  \bibinfo {author} {\bibfnamefont {C.~T.}\ \bibnamefont {Schmiegelow}},
  \bibinfo {author} {\bibfnamefont {V.}~\bibnamefont {Kaushal}}, \bibinfo
  {author} {\bibfnamefont {J.}~\bibnamefont {Schulz}}, \bibinfo {author}
  {\bibfnamefont {M.~T.}\ \bibnamefont {Mitchison}}, \bibinfo {author}
  {\bibfnamefont {J.}~\bibnamefont {Goold}}, \bibinfo {author} {\bibfnamefont
  {F.}~\bibnamefont {Schmidt-Kaler}}, \ and\ \bibinfo {author} {\bibfnamefont
  {U.~G.}\ \bibnamefont {Poschinger}},\ }\href {\doibase
  10.1103/PhysRevLett.123.080602} {\bibfield  {journal} {\bibinfo  {journal}
  {Phys. Rev. Lett.}\ }\textbf {\bibinfo {volume} {123}},\ \bibinfo {pages}
  {080602} (\bibinfo {year} {2019})}\BibitemShut {NoStop}%
\bibitem [{\citenamefont {Klatzow}\ \emph {et~al.}(2019)\citenamefont
  {Klatzow}, \citenamefont {Becker}, \citenamefont {Ledingham}, \citenamefont
  {Weinzetl}, \citenamefont {Kaczmarek}, \citenamefont {Saunders},
  \citenamefont {Nunn}, \citenamefont {Walmsley}, \citenamefont {Uzdin},\ and\
  \citenamefont {Poem}}]{KlatzowPRL2019}%
  \BibitemOpen
  \bibfield  {author} {\bibinfo {author} {\bibfnamefont {J.}~\bibnamefont
  {Klatzow}}, \bibinfo {author} {\bibfnamefont {J.~N.}\ \bibnamefont {Becker}},
  \bibinfo {author} {\bibfnamefont {P.~M.}\ \bibnamefont {Ledingham}}, \bibinfo
  {author} {\bibfnamefont {C.}~\bibnamefont {Weinzetl}}, \bibinfo {author}
  {\bibfnamefont {K.~T.}\ \bibnamefont {Kaczmarek}}, \bibinfo {author}
  {\bibfnamefont {D.~J.}\ \bibnamefont {Saunders}}, \bibinfo {author}
  {\bibfnamefont {J.}~\bibnamefont {Nunn}}, \bibinfo {author} {\bibfnamefont
  {I.~A.}\ \bibnamefont {Walmsley}}, \bibinfo {author} {\bibfnamefont
  {R.}~\bibnamefont {Uzdin}}, \ and\ \bibinfo {author} {\bibfnamefont
  {E.}~\bibnamefont {Poem}},\ }\href {\doibase 10.1103/PhysRevLett.122.110601}
  {\bibfield  {journal} {\bibinfo  {journal} {Phys. Rev. Lett.}\ }\textbf
  {\bibinfo {volume} {122}},\ \bibinfo {pages} {110601} (\bibinfo {year}
  {2019})}\BibitemShut {NoStop}%
\bibitem [{\citenamefont {Peterson}\ \emph {et~al.}(2019)\citenamefont
  {Peterson}, \citenamefont {Batalh\~ao}, \citenamefont {Herrera},
  \citenamefont {Souza}, \citenamefont {Sarthour}, \citenamefont {Oliveira},\
  and\ \citenamefont {Serra}}]{PetersonPRL2019}%
  \BibitemOpen
  \bibfield  {author} {\bibinfo {author} {\bibfnamefont {J.~P.~S.}\
  \bibnamefont {Peterson}}, \bibinfo {author} {\bibfnamefont {T.~B.}\
  \bibnamefont {Batalh\~ao}}, \bibinfo {author} {\bibfnamefont
  {M.}~\bibnamefont {Herrera}}, \bibinfo {author} {\bibfnamefont {A.~M.}\
  \bibnamefont {Souza}}, \bibinfo {author} {\bibfnamefont {R.~S.}\ \bibnamefont
  {Sarthour}}, \bibinfo {author} {\bibfnamefont {I.~S.}\ \bibnamefont
  {Oliveira}}, \ and\ \bibinfo {author} {\bibfnamefont {R.~M.}\ \bibnamefont
  {Serra}},\ }\href {\doibase 10.1103/PhysRevLett.123.240601} {\bibfield
  {journal} {\bibinfo  {journal} {Phys. Rev. Lett.}\ }\textbf {\bibinfo
  {volume} {123}},\ \bibinfo {pages} {240601} (\bibinfo {year}
  {2019})}\BibitemShut {NoStop}%
\bibitem [{\citenamefont {Gluza}\ \emph {et~al.}(2020)\citenamefont {Gluza},
  \citenamefont {Sabino}, \citenamefont {Ng}, \citenamefont {Vitagliano},
  \citenamefont {Pezzutto}, \citenamefont {Omar}, \citenamefont {Mazets},
  \citenamefont {Huber}, \citenamefont {Schmiedmayer},\ and\ \citenamefont
  {Eisert}}]{gluza2020}%
  \BibitemOpen
  \bibfield  {author} {\bibinfo {author} {\bibfnamefont {M.}~\bibnamefont
  {Gluza}}, \bibinfo {author} {\bibfnamefont {J.}~\bibnamefont {Sabino}},
  \bibinfo {author} {\bibfnamefont {N.~H.~Y.}\ \bibnamefont {Ng}}, \bibinfo
  {author} {\bibfnamefont {G.}~\bibnamefont {Vitagliano}}, \bibinfo {author}
  {\bibfnamefont {M.}~\bibnamefont {Pezzutto}}, \bibinfo {author}
  {\bibfnamefont {Y.}~\bibnamefont {Omar}}, \bibinfo {author} {\bibfnamefont
  {I.}~\bibnamefont {Mazets}}, \bibinfo {author} {\bibfnamefont
  {M.}~\bibnamefont {Huber}}, \bibinfo {author} {\bibfnamefont
  {J.}~\bibnamefont {Schmiedmayer}}, \ and\ \bibinfo {author} {\bibfnamefont
  {J.}~\bibnamefont {Eisert}},\ }\href@noop {} {\enquote {\bibinfo {title}
  {Quantum field thermal machines},}\ } (\bibinfo {year} {2020}),\ \Eprint
  {http://arxiv.org/abs/2006.01177} {arXiv:2006.01177 [quant-ph]} \BibitemShut
  {NoStop}%
\bibitem [{\citenamefont {Leggio}\ \emph {et~al.}(2015)\citenamefont {Leggio},
  \citenamefont {Bellomo},\ and\ \citenamefont {Antezza}}]{LeggioPRA2015}%
  \BibitemOpen
  \bibfield  {author} {\bibinfo {author} {\bibfnamefont {B.}~\bibnamefont
  {Leggio}}, \bibinfo {author} {\bibfnamefont {B.}~\bibnamefont {Bellomo}}, \
  and\ \bibinfo {author} {\bibfnamefont {M.}~\bibnamefont {Antezza}},\ }\href
  {\doibase 10.1103/PhysRevA.91.012117} {\bibfield  {journal} {\bibinfo
  {journal} {Phys. Rev. A}\ }\textbf {\bibinfo {volume} {91}},\ \bibinfo
  {pages} {012117} (\bibinfo {year} {2015})}\BibitemShut {NoStop}%
\bibitem [{\citenamefont {Leggio}\ and\ \citenamefont
  {Antezza}(2016)}]{LeggioPRE2016}%
  \BibitemOpen
  \bibfield  {author} {\bibinfo {author} {\bibfnamefont {B.}~\bibnamefont
  {Leggio}}\ and\ \bibinfo {author} {\bibfnamefont {M.}~\bibnamefont
  {Antezza}},\ }\href {\doibase 10.1103/PhysRevE.93.022122} {\bibfield
  {journal} {\bibinfo  {journal} {Phys. Rev. E}\ }\textbf {\bibinfo {volume}
  {93}},\ \bibinfo {pages} {022122} (\bibinfo {year} {2016})}\BibitemShut
  {NoStop}%
\bibitem [{\citenamefont {Reid}\ \emph {et~al.}(2017)\citenamefont {Reid},
  \citenamefont {Pigeon}, \citenamefont {Antezza},\ and\ \citenamefont {{De
  Chiara}}}]{ReidEPL2017}%
  \BibitemOpen
  \bibfield  {author} {\bibinfo {author} {\bibfnamefont {B.}~\bibnamefont
  {Reid}}, \bibinfo {author} {\bibfnamefont {S.}~\bibnamefont {Pigeon}},
  \bibinfo {author} {\bibfnamefont {M.}~\bibnamefont {Antezza}}, \ and\
  \bibinfo {author} {\bibfnamefont {G.}~\bibnamefont {{De Chiara}}},\ }\href
  {\doibase 10.1209/0295-5075/120/60006} {\bibfield  {journal} {\bibinfo
  {journal} {{EPL} (Europhysics Letters)}\ }\textbf {\bibinfo {volume} {120}},\
  \bibinfo {pages} {60006} (\bibinfo {year} {2017})}\BibitemShut {NoStop}%
\bibitem [{\citenamefont {de~Assis}\ \emph {et~al.}(2019)\citenamefont
  {de~Assis}, \citenamefont {de~Mendon\c{c}a}, \citenamefont {Villas-Boas},
  \citenamefont {de~Souza}, \citenamefont {Sarthour}, \citenamefont
  {Oliveira},\ and\ \citenamefont {de~Almeida}}]{AssisPRL2019}%
  \BibitemOpen
  \bibfield  {author} {\bibinfo {author} {\bibfnamefont {R.~J.}\ \bibnamefont
  {de~Assis}}, \bibinfo {author} {\bibfnamefont {T.~M.}\ \bibnamefont
  {de~Mendon\c{c}a}}, \bibinfo {author} {\bibfnamefont {C.~J.}\ \bibnamefont
  {Villas-Boas}}, \bibinfo {author} {\bibfnamefont {A.~M.}\ \bibnamefont
  {de~Souza}}, \bibinfo {author} {\bibfnamefont {R.~S.}\ \bibnamefont
  {Sarthour}}, \bibinfo {author} {\bibfnamefont {I.~S.}\ \bibnamefont
  {Oliveira}}, \ and\ \bibinfo {author} {\bibfnamefont {N.~G.}\ \bibnamefont
  {de~Almeida}},\ }\href {\doibase 10.1103/PhysRevLett.122.240602} {\bibfield
  {journal} {\bibinfo  {journal} {Phys. Rev. Lett.}\ }\textbf {\bibinfo
  {volume} {122}},\ \bibinfo {pages} {240602} (\bibinfo {year}
  {2019})}\BibitemShut {NoStop}%
\bibitem [{\citenamefont {Cherubim}\ \emph {et~al.}(2019)\citenamefont
  {Cherubim}, \citenamefont {Brito},\ and\ \citenamefont
  {Deffner}}]{cherubim2019}%
  \BibitemOpen
  \bibfield  {author} {\bibinfo {author} {\bibfnamefont {C.}~\bibnamefont
  {Cherubim}}, \bibinfo {author} {\bibfnamefont {F.}~\bibnamefont {Brito}}, \
  and\ \bibinfo {author} {\bibfnamefont {S.}~\bibnamefont {Deffner}},\
  }\href@noop {} {\bibfield  {journal} {\bibinfo  {journal} {Entropy}\ }\textbf
  {\bibinfo {volume} {21}},\ \bibinfo {pages} {545} (\bibinfo {year}
  {2019})}\BibitemShut {NoStop}%
\bibitem [{\citenamefont {Pezzutto}\ \emph {et~al.}(2019)\citenamefont
  {Pezzutto}, \citenamefont {Paternostro},\ and\ \citenamefont
  {Omar}}]{PezzuttoQST2019}%
  \BibitemOpen
  \bibfield  {author} {\bibinfo {author} {\bibfnamefont {M.}~\bibnamefont
  {Pezzutto}}, \bibinfo {author} {\bibfnamefont {M.}~\bibnamefont
  {Paternostro}}, \ and\ \bibinfo {author} {\bibfnamefont {Y.}~\bibnamefont
  {Omar}},\ }\href {\doibase 10.1088/2058-9565/aaf5b4} {\bibfield  {journal}
  {\bibinfo  {journal} {Quantum Science and Technology}\ }\textbf {\bibinfo
  {volume} {4}},\ \bibinfo {pages} {025002} (\bibinfo {year}
  {2019})}\BibitemShut {NoStop}%
\bibitem [{\citenamefont {Carollo}\ \emph {et~al.}(2020)\citenamefont
  {Carollo}, \citenamefont {Gambetta}, \citenamefont {Brandner}, \citenamefont
  {Garrahan},\ and\ \citenamefont {Lesanovsky}}]{CarolloPRL2020}%
  \BibitemOpen
  \bibfield  {author} {\bibinfo {author} {\bibfnamefont {F.}~\bibnamefont
  {Carollo}}, \bibinfo {author} {\bibfnamefont {F.~M.}\ \bibnamefont
  {Gambetta}}, \bibinfo {author} {\bibfnamefont {K.}~\bibnamefont {Brandner}},
  \bibinfo {author} {\bibfnamefont {J.~P.}\ \bibnamefont {Garrahan}}, \ and\
  \bibinfo {author} {\bibfnamefont {I.}~\bibnamefont {Lesanovsky}},\ }\href
  {\doibase 10.1103/PhysRevLett.124.170602} {\bibfield  {journal} {\bibinfo
  {journal} {Phys. Rev. Lett.}\ }\textbf {\bibinfo {volume} {124}},\ \bibinfo
  {pages} {170602} (\bibinfo {year} {2020})}\BibitemShut {NoStop}%
\bibitem [{\citenamefont {Rom{\'{a}}n-Ancheyta}\ \emph
  {et~al.}(2019)\citenamefont {Rom{\'{a}}n-Ancheyta}, \citenamefont
  {{\c{C}}akmak},\ and\ \citenamefont
  {M\"ustecapl{\i}o\u{g}lu}}]{AncheytaQST2019}%
  \BibitemOpen
  \bibfield  {author} {\bibinfo {author} {\bibfnamefont {R.}~\bibnamefont
  {Rom{\'{a}}n-Ancheyta}}, \bibinfo {author} {\bibfnamefont {B.}~\bibnamefont
  {{\c{C}}akmak}}, \ and\ \bibinfo {author} {\bibfnamefont {{\"O}.~E.}\
  \bibnamefont {M\"ustecapl{\i}o\u{g}lu}},\ }\href {\doibase
  10.1088/2058-9565/ab5e4f} {\bibfield  {journal} {\bibinfo  {journal} {Quantum
  Science and Technology}\ }\textbf {\bibinfo {volume} {5}},\ \bibinfo {pages}
  {015003} (\bibinfo {year} {2019})}\BibitemShut {NoStop}%
\bibitem [{\citenamefont {Ro\ss{}nagel}\ \emph {et~al.}(2014)\citenamefont
  {Ro\ss{}nagel}, \citenamefont {Abah}, \citenamefont {Schmidt-Kaler},
  \citenamefont {Singer},\ and\ \citenamefont {Lutz}}]{RossnagelPRL2014}%
  \BibitemOpen
  \bibfield  {author} {\bibinfo {author} {\bibfnamefont {J.}~\bibnamefont
  {Ro\ss{}nagel}}, \bibinfo {author} {\bibfnamefont {O.}~\bibnamefont {Abah}},
  \bibinfo {author} {\bibfnamefont {F.}~\bibnamefont {Schmidt-Kaler}}, \bibinfo
  {author} {\bibfnamefont {K.}~\bibnamefont {Singer}}, \ and\ \bibinfo {author}
  {\bibfnamefont {E.}~\bibnamefont {Lutz}},\ }\href {\doibase
  10.1103/PhysRevLett.112.030602} {\bibfield  {journal} {\bibinfo  {journal}
  {Phys. Rev. Lett.}\ }\textbf {\bibinfo {volume} {112}},\ \bibinfo {pages}
  {030602} (\bibinfo {year} {2014})}\BibitemShut {NoStop}%
\bibitem [{\citenamefont {Manzano}\ \emph {et~al.}(2016)\citenamefont
  {Manzano}, \citenamefont {Galve}, \citenamefont {Zambrini},\ and\
  \citenamefont {Parrondo}}]{Manzano2016}%
  \BibitemOpen
  \bibfield  {author} {\bibinfo {author} {\bibfnamefont {G.}~\bibnamefont
  {Manzano}}, \bibinfo {author} {\bibfnamefont {F.}~\bibnamefont {Galve}},
  \bibinfo {author} {\bibfnamefont {R.}~\bibnamefont {Zambrini}}, \ and\
  \bibinfo {author} {\bibfnamefont {J.~M.~R.}\ \bibnamefont {Parrondo}},\
  }\href {\doibase 10.1103/PhysRevE.93.052120} {\bibfield  {journal} {\bibinfo
  {journal} {Phys. Rev. E}\ }\textbf {\bibinfo {volume} {93}},\ \bibinfo
  {pages} {052120} (\bibinfo {year} {2016})}\BibitemShut {NoStop}%
\bibitem [{\citenamefont {Agarwalla}\ \emph {et~al.}(2017)\citenamefont
  {Agarwalla}, \citenamefont {Jiang},\ and\ \citenamefont
  {Segal}}]{AgarwallaPRB2017}%
  \BibitemOpen
  \bibfield  {author} {\bibinfo {author} {\bibfnamefont {B.~K.}\ \bibnamefont
  {Agarwalla}}, \bibinfo {author} {\bibfnamefont {J.-H.}\ \bibnamefont
  {Jiang}}, \ and\ \bibinfo {author} {\bibfnamefont {D.}~\bibnamefont
  {Segal}},\ }\href {\doibase 10.1103/PhysRevB.96.104304} {\bibfield  {journal}
  {\bibinfo  {journal} {Phys. Rev. B}\ }\textbf {\bibinfo {volume} {96}},\
  \bibinfo {pages} {104304} (\bibinfo {year} {2017})}\BibitemShut {NoStop}%
\bibitem [{\citenamefont {Singh}\ and\ \citenamefont
  {M\"ustecapl{\i}o\u{g}lu}(2020)}]{singh2020performance}%
  \BibitemOpen
  \bibfield  {author} {\bibinfo {author} {\bibfnamefont {V.}~\bibnamefont
  {Singh}}\ and\ \bibinfo {author} {\bibfnamefont {{\"O}.~E.}\ \bibnamefont
  {M\"ustecapl{\i}o\u{g}lu}},\ }\href@noop {} {\enquote {\bibinfo {title}
  {Performance bounds of non-adiabatic quantum harmonic otto engine and
  refrigerator under a squeezed thermal reservoir},}\ } (\bibinfo {year}
  {2020}),\ \Eprint {http://arxiv.org/abs/2006.08311} {arXiv:2006.08311
  [quant-ph]} \BibitemShut {NoStop}%
\bibitem [{\citenamefont {Niedenzu}\ \emph {et~al.}(2018)\citenamefont
  {Niedenzu}, \citenamefont {Mukherjee}, \citenamefont {Ghosh}, \citenamefont
  {Kofman},\ and\ \citenamefont {Kurizki}}]{niedenzu2018quantum}%
  \BibitemOpen
  \bibfield  {author} {\bibinfo {author} {\bibfnamefont {W.}~\bibnamefont
  {Niedenzu}}, \bibinfo {author} {\bibfnamefont {V.}~\bibnamefont {Mukherjee}},
  \bibinfo {author} {\bibfnamefont {A.}~\bibnamefont {Ghosh}}, \bibinfo
  {author} {\bibfnamefont {A.~G.}\ \bibnamefont {Kofman}}, \ and\ \bibinfo
  {author} {\bibfnamefont {G.}~\bibnamefont {Kurizki}},\ }\href@noop {}
  {\bibfield  {journal} {\bibinfo  {journal} {Nature communications}\ }\textbf
  {\bibinfo {volume} {9}},\ \bibinfo {pages} {1} (\bibinfo {year}
  {2018})}\BibitemShut {NoStop}%
\bibitem [{\citenamefont {Doyeux}\ \emph {et~al.}(2016)\citenamefont {Doyeux},
  \citenamefont {Leggio}, \citenamefont {Messina},\ and\ \citenamefont
  {Antezza}}]{DoyeuxPRE2016}%
  \BibitemOpen
  \bibfield  {author} {\bibinfo {author} {\bibfnamefont {P.}~\bibnamefont
  {Doyeux}}, \bibinfo {author} {\bibfnamefont {B.}~\bibnamefont {Leggio}},
  \bibinfo {author} {\bibfnamefont {R.}~\bibnamefont {Messina}}, \ and\
  \bibinfo {author} {\bibfnamefont {M.}~\bibnamefont {Antezza}},\ }\href
  {\doibase 10.1103/PhysRevE.93.022134} {\bibfield  {journal} {\bibinfo
  {journal} {Phys. Rev. E}\ }\textbf {\bibinfo {volume} {93}},\ \bibinfo
  {pages} {022134} (\bibinfo {year} {2016})}\BibitemShut {NoStop}%
\bibitem [{\citenamefont {T\"urkpen{\c{c}}e}\ \emph {et~al.}(2017)\citenamefont
  {T\"urkpen{\c{c}}e}, \citenamefont {Altintas}, \citenamefont {Paternostro},\
  and\ \citenamefont {M\"ustecapl{\i}o\u{g}lu}}]{TurkpenceEPL2017}%
  \BibitemOpen
  \bibfield  {author} {\bibinfo {author} {\bibfnamefont {D.}~\bibnamefont
  {T\"urkpen{\c{c}}e}}, \bibinfo {author} {\bibfnamefont {F.}~\bibnamefont
  {Altintas}}, \bibinfo {author} {\bibfnamefont {M.}~\bibnamefont
  {Paternostro}}, \ and\ \bibinfo {author} {\bibfnamefont {{\"O}.~E.}\
  \bibnamefont {M\"ustecapl{\i}o\u{g}lu}},\ }\href {\doibase
  10.1209/0295-5075/117/50002} {\bibfield  {journal} {\bibinfo  {journal}
  {{EPL} (Europhysics Letters)}\ }\textbf {\bibinfo {volume} {117}},\ \bibinfo
  {pages} {50002} (\bibinfo {year} {2017})}\BibitemShut {NoStop}%
\bibitem [{\citenamefont {Karimi}\ and\ \citenamefont
  {Pekola}(2017)}]{KarimiPRB2017}%
  \BibitemOpen
  \bibfield  {author} {\bibinfo {author} {\bibfnamefont {B.}~\bibnamefont
  {Karimi}}\ and\ \bibinfo {author} {\bibfnamefont {J.~P.}\ \bibnamefont
  {Pekola}},\ }\href {\doibase 10.1103/PhysRevB.96.115408} {\bibfield
  {journal} {\bibinfo  {journal} {Phys. Rev. B}\ }\textbf {\bibinfo {volume}
  {96}},\ \bibinfo {pages} {115408} (\bibinfo {year} {2017})}\BibitemShut
  {NoStop}%
\bibitem [{\citenamefont {Hewgill}\ \emph {et~al.}(2018)\citenamefont
  {Hewgill}, \citenamefont {Ferraro},\ and\ \citenamefont {{De
  Chiara}}}]{HewgillPRA2018}%
  \BibitemOpen
  \bibfield  {author} {\bibinfo {author} {\bibfnamefont {A.}~\bibnamefont
  {Hewgill}}, \bibinfo {author} {\bibfnamefont {A.}~\bibnamefont {Ferraro}}, \
  and\ \bibinfo {author} {\bibfnamefont {G.}~\bibnamefont {{De Chiara}}},\
  }\href {\doibase 10.1103/PhysRevA.98.042102} {\bibfield  {journal} {\bibinfo
  {journal} {Phys. Rev. A}\ }\textbf {\bibinfo {volume} {98}},\ \bibinfo
  {pages} {042102} (\bibinfo {year} {2018})}\BibitemShut {NoStop}%
\bibitem [{\citenamefont {Latune}\ \emph {et~al.}(2019)\citenamefont {Latune},
  \citenamefont {Sinayskiy},\ and\ \citenamefont
  {Petruccione}}]{LatuneQST2019}%
  \BibitemOpen
  \bibfield  {author} {\bibinfo {author} {\bibfnamefont {C.~L.}\ \bibnamefont
  {Latune}}, \bibinfo {author} {\bibfnamefont {I.}~\bibnamefont {Sinayskiy}}, \
  and\ \bibinfo {author} {\bibfnamefont {F.}~\bibnamefont {Petruccione}},\
  }\href {\doibase 10.1088/2058-9565/aaf5f7} {\bibfield  {journal} {\bibinfo
  {journal} {Quantum Science and Technology}\ }\textbf {\bibinfo {volume}
  {4}},\ \bibinfo {pages} {025005} (\bibinfo {year} {2019})}\BibitemShut
  {NoStop}%
\bibitem [{\citenamefont {Manzano}\ \emph {et~al.}(2019)\citenamefont
  {Manzano}, \citenamefont {Giorgi}, \citenamefont {Fazio},\ and\ \citenamefont
  {Zambrini}}]{ManzanoNJP2019}%
  \BibitemOpen
  \bibfield  {author} {\bibinfo {author} {\bibfnamefont {G.}~\bibnamefont
  {Manzano}}, \bibinfo {author} {\bibfnamefont {G.-L.}\ \bibnamefont {Giorgi}},
  \bibinfo {author} {\bibfnamefont {R.}~\bibnamefont {Fazio}}, \ and\ \bibinfo
  {author} {\bibfnamefont {R.}~\bibnamefont {Zambrini}},\ }\href {\doibase
  10.1088/1367-2630/ab5c58} {\bibfield  {journal} {\bibinfo  {journal} {New
  Journal of Physics}\ }\textbf {\bibinfo {volume} {21}},\ \bibinfo {pages}
  {123026} (\bibinfo {year} {2019})}\BibitemShut {NoStop}%
\bibitem [{\citenamefont {Pusuluk}\ and\ \citenamefont
  {M\"ustecapl{\i}o\u{g}lu}(2020)}]{pusuluk2020}%
  \BibitemOpen
  \bibfield  {author} {\bibinfo {author} {\bibfnamefont {O.}~\bibnamefont
  {Pusuluk}}\ and\ \bibinfo {author} {\bibfnamefont {{\"O}.~E.}\ \bibnamefont
  {M\"ustecapl{\i}o\u{g}lu}},\ }\href@noop {} {\enquote {\bibinfo {title}
  {Thermocoherent effect: heat currents driven by quantum coherence and
  correlations},}\ } (\bibinfo {year} {2020}),\ \Eprint
  {http://arxiv.org/abs/2006.03186} {arXiv:2006.03186 [quant-ph]} \BibitemShut
  {NoStop}%
\bibitem [{\citenamefont {Scarani}\ \emph {et~al.}(2002)\citenamefont
  {Scarani}, \citenamefont {Ziman}, \citenamefont {\ifmmode \check{S}\else
  \v{S}\fi{}telmachovi\ifmmode~\check{c}\else \v{c}\fi{}}, \citenamefont
  {Gisin},\ and\ \citenamefont {Bu\ifmmode~\check{z}\else
  \v{z}\fi{}ek}}]{ScaraniPRL2002}%
  \BibitemOpen
  \bibfield  {author} {\bibinfo {author} {\bibfnamefont {V.}~\bibnamefont
  {Scarani}}, \bibinfo {author} {\bibfnamefont {M.}~\bibnamefont {Ziman}},
  \bibinfo {author} {\bibfnamefont {P.}~\bibnamefont {\ifmmode \check{S}\else
  \v{S}\fi{}telmachovi\ifmmode~\check{c}\else \v{c}\fi{}}}, \bibinfo {author}
  {\bibfnamefont {N.}~\bibnamefont {Gisin}}, \ and\ \bibinfo {author}
  {\bibfnamefont {V.}~\bibnamefont {Bu\ifmmode~\check{z}\else \v{z}\fi{}ek}},\
  }\href {\doibase 10.1103/PhysRevLett.88.097905} {\bibfield  {journal}
  {\bibinfo  {journal} {Phys. Rev. Lett.}\ }\textbf {\bibinfo {volume} {88}},\
  \bibinfo {pages} {097905} (\bibinfo {year} {2002})}\BibitemShut {NoStop}%
\bibitem [{\citenamefont {Ziman}\ \emph {et~al.}(2002)\citenamefont {Ziman},
  \citenamefont {\ifmmode \check{S}\else
  \v{S}\fi{}telmachovi\ifmmode~\check{c}\else \v{c}\fi{}}, \citenamefont
  {Bu\ifmmode~\check{z}\else \v{z}\fi{}ek}, \citenamefont {Hillery},
  \citenamefont {Scarani},\ and\ \citenamefont {Gisin}}]{ZimanPRA2002}%
  \BibitemOpen
  \bibfield  {author} {\bibinfo {author} {\bibfnamefont {M.}~\bibnamefont
  {Ziman}}, \bibinfo {author} {\bibfnamefont {P.}~\bibnamefont {\ifmmode
  \check{S}\else \v{S}\fi{}telmachovi\ifmmode~\check{c}\else \v{c}\fi{}}},
  \bibinfo {author} {\bibfnamefont {V.}~\bibnamefont {Bu\ifmmode~\check{z}\else
  \v{z}\fi{}ek}}, \bibinfo {author} {\bibfnamefont {M.}~\bibnamefont
  {Hillery}}, \bibinfo {author} {\bibfnamefont {V.}~\bibnamefont {Scarani}}, \
  and\ \bibinfo {author} {\bibfnamefont {N.}~\bibnamefont {Gisin}},\ }\href
  {\doibase 10.1103/PhysRevA.65.042105} {\bibfield  {journal} {\bibinfo
  {journal} {Phys. Rev. A}\ }\textbf {\bibinfo {volume} {65}},\ \bibinfo
  {pages} {042105} (\bibinfo {year} {2002})}\BibitemShut {NoStop}%
\bibitem [{\citenamefont {Karevski}\ and\ \citenamefont
  {Platini}(2009)}]{Karevski2009}%
  \BibitemOpen
  \bibfield  {author} {\bibinfo {author} {\bibfnamefont {D.}~\bibnamefont
  {Karevski}}\ and\ \bibinfo {author} {\bibfnamefont {T.}~\bibnamefont
  {Platini}},\ }\href {\doibase 10.1103/PhysRevLett.102.207207} {\bibfield
  {journal} {\bibinfo  {journal} {Phys. Rev. Lett.}\ }\textbf {\bibinfo
  {volume} {102}},\ \bibinfo {pages} {207207} (\bibinfo {year}
  {2009})}\BibitemShut {NoStop}%
\bibitem [{\citenamefont {Giovannetti}\ and\ \citenamefont
  {Palma}(2012)}]{PalmaGiovannetti}%
  \BibitemOpen
  \bibfield  {author} {\bibinfo {author} {\bibfnamefont {V.}~\bibnamefont
  {Giovannetti}}\ and\ \bibinfo {author} {\bibfnamefont {G.~M.}\ \bibnamefont
  {Palma}},\ }\href {\doibase 10.1103/PhysRevLett.108.040401} {\bibfield
  {journal} {\bibinfo  {journal} {Phys. Rev. Lett.}\ }\textbf {\bibinfo
  {volume} {108}},\ \bibinfo {pages} {040401} (\bibinfo {year}
  {2012})}\BibitemShut {NoStop}%
\bibitem [{\citenamefont {Ciccarello}\ \emph {et~al.}(2013)\citenamefont
  {Ciccarello}, \citenamefont {Palma},\ and\ \citenamefont
  {Giovannetti}}]{CiccarelloPRA2013}%
  \BibitemOpen
  \bibfield  {author} {\bibinfo {author} {\bibfnamefont {F.}~\bibnamefont
  {Ciccarello}}, \bibinfo {author} {\bibfnamefont {G.~M.}\ \bibnamefont
  {Palma}}, \ and\ \bibinfo {author} {\bibfnamefont {V.}~\bibnamefont
  {Giovannetti}},\ }\href {\doibase 10.1103/PhysRevA.87.040103} {\bibfield
  {journal} {\bibinfo  {journal} {Phys. Rev. A}\ }\textbf {\bibinfo {volume}
  {87}},\ \bibinfo {pages} {040103(R)} (\bibinfo {year} {2013})}\BibitemShut
  {NoStop}%
\bibitem [{\citenamefont {Lorenzo}\ \emph
  {et~al.}(2015{\natexlab{a}})\citenamefont {Lorenzo}, \citenamefont {Farace},
  \citenamefont {Ciccarello}, \citenamefont {Palma},\ and\ \citenamefont
  {Giovannetti}}]{LorenzoPRA2015}%
  \BibitemOpen
  \bibfield  {author} {\bibinfo {author} {\bibfnamefont {S.}~\bibnamefont
  {Lorenzo}}, \bibinfo {author} {\bibfnamefont {A.}~\bibnamefont {Farace}},
  \bibinfo {author} {\bibfnamefont {F.}~\bibnamefont {Ciccarello}}, \bibinfo
  {author} {\bibfnamefont {G.~M.}\ \bibnamefont {Palma}}, \ and\ \bibinfo
  {author} {\bibfnamefont {V.}~\bibnamefont {Giovannetti}},\ }\href {\doibase
  10.1103/PhysRevA.91.022121} {\bibfield  {journal} {\bibinfo  {journal} {Phys.
  Rev. A}\ }\textbf {\bibinfo {volume} {91}},\ \bibinfo {pages} {022121}
  (\bibinfo {year} {2015}{\natexlab{a}})}\BibitemShut {NoStop}%
\bibitem [{\citenamefont {Landi}\ \emph {et~al.}(2014)\citenamefont {Landi},
  \citenamefont {Novais}, \citenamefont {de~Oliveira},\ and\ \citenamefont
  {Karevski}}]{Landi2014b}%
  \BibitemOpen
  \bibfield  {author} {\bibinfo {author} {\bibfnamefont {G.~T.}\ \bibnamefont
  {Landi}}, \bibinfo {author} {\bibfnamefont {E.}~\bibnamefont {Novais}},
  \bibinfo {author} {\bibfnamefont {M.~J.}\ \bibnamefont {de~Oliveira}}, \ and\
  \bibinfo {author} {\bibfnamefont {D.}~\bibnamefont {Karevski}},\ }\href
  {\doibase 10.1103/PhysRevE.90.042142} {\bibfield  {journal} {\bibinfo
  {journal} {Phys. Rev. E}\ }\textbf {\bibinfo {volume} {90}},\ \bibinfo
  {pages} {042142} (\bibinfo {year} {2014})}\BibitemShut {NoStop}%
\bibitem [{\citenamefont {Lorenzo}\ \emph
  {et~al.}(2015{\natexlab{b}})\citenamefont {Lorenzo}, \citenamefont
  {McCloskey}, \citenamefont {Ciccarello}, \citenamefont {Paternostro},\ and\
  \citenamefont {Palma}}]{LorenzoPRL2015}%
  \BibitemOpen
  \bibfield  {author} {\bibinfo {author} {\bibfnamefont {S.}~\bibnamefont
  {Lorenzo}}, \bibinfo {author} {\bibfnamefont {R.}~\bibnamefont {McCloskey}},
  \bibinfo {author} {\bibfnamefont {F.}~\bibnamefont {Ciccarello}}, \bibinfo
  {author} {\bibfnamefont {M.}~\bibnamefont {Paternostro}}, \ and\ \bibinfo
  {author} {\bibfnamefont {G.~M.}\ \bibnamefont {Palma}},\ }\href {\doibase
  10.1103/PhysRevLett.115.120403} {\bibfield  {journal} {\bibinfo  {journal}
  {Phys. Rev. Lett.}\ }\textbf {\bibinfo {volume} {115}},\ \bibinfo {pages}
  {120403} (\bibinfo {year} {2015}{\natexlab{b}})}\BibitemShut {NoStop}%
\bibitem [{\citenamefont {Barra}(2015)}]{Barra2015}%
  \BibitemOpen
  \bibfield  {author} {\bibinfo {author} {\bibfnamefont {F.}~\bibnamefont
  {Barra}},\ }\href {\doibase 10.1038/srep14873} {\bibfield  {journal}
  {\bibinfo  {journal} {Scientific Reports}\ }\textbf {\bibinfo {volume} {5}},\
  \bibinfo {pages} {14873} (\bibinfo {year} {2015})}\BibitemShut {NoStop}%
\bibitem [{\citenamefont {Strasberg}\ \emph {et~al.}(2017)\citenamefont
  {Strasberg}, \citenamefont {Schaller}, \citenamefont {Brandes},\ and\
  \citenamefont {Esposito}}]{Strasberg2016}%
  \BibitemOpen
  \bibfield  {author} {\bibinfo {author} {\bibfnamefont {P.}~\bibnamefont
  {Strasberg}}, \bibinfo {author} {\bibfnamefont {G.}~\bibnamefont {Schaller}},
  \bibinfo {author} {\bibfnamefont {T.}~\bibnamefont {Brandes}}, \ and\
  \bibinfo {author} {\bibfnamefont {M.}~\bibnamefont {Esposito}},\ }\href
  {\doibase 10.1103/PhysRevX.7.021003} {\bibfield  {journal} {\bibinfo
  {journal} {Phys. Rev. X}\ }\textbf {\bibinfo {volume} {7}},\ \bibinfo {pages}
  {021003} (\bibinfo {year} {2017})}\BibitemShut {NoStop}%
\bibitem [{\citenamefont {Ciccarello}(2017)}]{Ciccarello2017}%
  \BibitemOpen
  \bibfield  {author} {\bibinfo {author} {\bibfnamefont {F.}~\bibnamefont
  {Ciccarello}},\ }\href {\doibase 10.1515/qmetro-2017-0007} {\bibfield
  {journal} {\bibinfo  {journal} {Quantum Measurements and Quantum Metrology}\
  }\textbf {\bibinfo {volume} {4}},\ \bibinfo {pages} {53} (\bibinfo {year}
  {2017})}\BibitemShut {NoStop}%
\bibitem [{\citenamefont {Pezzutto}\ \emph {et~al.}(2016)\citenamefont
  {Pezzutto}, \citenamefont {Paternostro},\ and\ \citenamefont
  {Omar}}]{PezzuttoNJP2016}%
  \BibitemOpen
  \bibfield  {author} {\bibinfo {author} {\bibfnamefont {M.}~\bibnamefont
  {Pezzutto}}, \bibinfo {author} {\bibfnamefont {M.}~\bibnamefont
  {Paternostro}}, \ and\ \bibinfo {author} {\bibfnamefont {Y.}~\bibnamefont
  {Omar}},\ }\href {http://stacks.iop.org/1367-2630/18/i=12/a=123018}
  {\bibfield  {journal} {\bibinfo  {journal} {New Journal of Physics}\ }\textbf
  {\bibinfo {volume} {18}},\ \bibinfo {pages} {123018} (\bibinfo {year}
  {2016})}\BibitemShut {NoStop}%
\bibitem [{\citenamefont {Cusumano}\ \emph {et~al.}(2018)\citenamefont
  {Cusumano}, \citenamefont {Cavina}, \citenamefont {Keck}, \citenamefont
  {De~Pasquale},\ and\ \citenamefont {Giovannetti}}]{cusumano2018entropy}%
  \BibitemOpen
  \bibfield  {author} {\bibinfo {author} {\bibfnamefont {S.}~\bibnamefont
  {Cusumano}}, \bibinfo {author} {\bibfnamefont {V.}~\bibnamefont {Cavina}},
  \bibinfo {author} {\bibfnamefont {M.}~\bibnamefont {Keck}}, \bibinfo {author}
  {\bibfnamefont {A.}~\bibnamefont {De~Pasquale}}, \ and\ \bibinfo {author}
  {\bibfnamefont {V.}~\bibnamefont {Giovannetti}},\ }\href {\doibase
  10.1103/PhysRevA.98.032119} {\bibfield  {journal} {\bibinfo  {journal} {Phys.
  Rev. A}\ }\textbf {\bibinfo {volume} {98}},\ \bibinfo {pages} {032119}
  (\bibinfo {year} {2018})}\BibitemShut {NoStop}%
\bibitem [{\citenamefont {Campbell}\ \emph {et~al.}(2018)\citenamefont
  {Campbell}, \citenamefont {Ciccarello}, \citenamefont {Palma},\ and\
  \citenamefont {Vacchini}}]{CampbellPRA2018}%
  \BibitemOpen
  \bibfield  {author} {\bibinfo {author} {\bibfnamefont {S.}~\bibnamefont
  {Campbell}}, \bibinfo {author} {\bibfnamefont {F.}~\bibnamefont
  {Ciccarello}}, \bibinfo {author} {\bibfnamefont {G.~M.}\ \bibnamefont
  {Palma}}, \ and\ \bibinfo {author} {\bibfnamefont {B.}~\bibnamefont
  {Vacchini}},\ }\href {\doibase 10.1103/PhysRevA.98.012142} {\bibfield
  {journal} {\bibinfo  {journal} {Phys. Rev. A}\ }\textbf {\bibinfo {volume}
  {98}},\ \bibinfo {pages} {012142} (\bibinfo {year} {2018})}\BibitemShut
  {NoStop}%
\bibitem [{\citenamefont {Gross}\ \emph {et~al.}(2018)\citenamefont {Gross},
  \citenamefont {Caves}, \citenamefont {Milburn},\ and\ \citenamefont
  {Combes}}]{GrossQST2018}%
  \BibitemOpen
  \bibfield  {author} {\bibinfo {author} {\bibfnamefont {J.~A.}\ \bibnamefont
  {Gross}}, \bibinfo {author} {\bibfnamefont {C.~M.}\ \bibnamefont {Caves}},
  \bibinfo {author} {\bibfnamefont {G.~J.}\ \bibnamefont {Milburn}}, \ and\
  \bibinfo {author} {\bibfnamefont {J.}~\bibnamefont {Combes}},\ }\href
  {http://stacks.iop.org/2058-9565/3/i=2/a=024005} {\bibfield  {journal}
  {\bibinfo  {journal} {Quantum Science and Technology}\ }\textbf {\bibinfo
  {volume} {3}},\ \bibinfo {pages} {024005} (\bibinfo {year}
  {2018})}\BibitemShut {NoStop}%
\bibitem [{\citenamefont {Mohammady}\ \emph {et~al.}(2018)\citenamefont
  {Mohammady}, \citenamefont {Choi}, \citenamefont {Trusheim}, \citenamefont
  {Bayat}, \citenamefont {Englund},\ and\ \citenamefont
  {Omar}}]{MohammadyPRA2018}%
  \BibitemOpen
  \bibfield  {author} {\bibinfo {author} {\bibfnamefont {M.~H.}\ \bibnamefont
  {Mohammady}}, \bibinfo {author} {\bibfnamefont {H.}~\bibnamefont {Choi}},
  \bibinfo {author} {\bibfnamefont {M.~E.}\ \bibnamefont {Trusheim}}, \bibinfo
  {author} {\bibfnamefont {A.}~\bibnamefont {Bayat}}, \bibinfo {author}
  {\bibfnamefont {D.}~\bibnamefont {Englund}}, \ and\ \bibinfo {author}
  {\bibfnamefont {Y.}~\bibnamefont {Omar}},\ }\href {\doibase
  10.1103/PhysRevA.97.042124} {\bibfield  {journal} {\bibinfo  {journal} {Phys.
  Rev. A}\ }\textbf {\bibinfo {volume} {97}},\ \bibinfo {pages} {042124}
  (\bibinfo {year} {2018})}\BibitemShut {NoStop}%
\bibitem [{\citenamefont {Rodrigues}\ \emph {et~al.}(2019)\citenamefont
  {Rodrigues}, \citenamefont {{De Chiara}}, \citenamefont {Paternostro},\ and\
  \citenamefont {Landi}}]{RodriguesPRL2019}%
  \BibitemOpen
  \bibfield  {author} {\bibinfo {author} {\bibfnamefont {F.~L.~S.}\
  \bibnamefont {Rodrigues}}, \bibinfo {author} {\bibfnamefont {G.}~\bibnamefont
  {{De Chiara}}}, \bibinfo {author} {\bibfnamefont {M.}~\bibnamefont
  {Paternostro}}, \ and\ \bibinfo {author} {\bibfnamefont {G.~T.}\ \bibnamefont
  {Landi}},\ }\href {\doibase 10.1103/PhysRevLett.123.140601} {\bibfield
  {journal} {\bibinfo  {journal} {Phys. Rev. Lett.}\ }\textbf {\bibinfo
  {volume} {123}},\ \bibinfo {pages} {140601} (\bibinfo {year}
  {2019})}\BibitemShut {NoStop}%
\bibitem [{\citenamefont {Manatuly}\ \emph {et~al.}(2019)\citenamefont
  {Manatuly}, \citenamefont {Niedenzu}, \citenamefont {Rom\'an-Ancheyta},
  \citenamefont {\c{C}akmak}, \citenamefont {M\"ustecapl{\i}o\u{g}lu},\ and\
  \citenamefont {Kurizki}}]{ManatulyPRE2019}%
  \BibitemOpen
  \bibfield  {author} {\bibinfo {author} {\bibfnamefont {A.}~\bibnamefont
  {Manatuly}}, \bibinfo {author} {\bibfnamefont {W.}~\bibnamefont {Niedenzu}},
  \bibinfo {author} {\bibfnamefont {R.}~\bibnamefont {Rom\'an-Ancheyta}},
  \bibinfo {author} {\bibfnamefont {B.}~\bibnamefont {\c{C}akmak}}, \bibinfo
  {author} {\bibfnamefont {{\"O}.~E.}\ \bibnamefont {M\"ustecapl{\i}o\u{g}lu}},
  \ and\ \bibinfo {author} {\bibfnamefont {G.}~\bibnamefont {Kurizki}},\ }\href
  {\doibase 10.1103/PhysRevE.99.042145} {\bibfield  {journal} {\bibinfo
  {journal} {Phys. Rev. E}\ }\textbf {\bibinfo {volume} {99}},\ \bibinfo
  {pages} {042145} (\bibinfo {year} {2019})}\BibitemShut {NoStop}%
\bibitem [{\citenamefont {Strasberg}(2019)}]{StrasbergPRL2019}%
  \BibitemOpen
  \bibfield  {author} {\bibinfo {author} {\bibfnamefont {P.}~\bibnamefont
  {Strasberg}},\ }\href {\doibase 10.1103/PhysRevLett.123.180604} {\bibfield
  {journal} {\bibinfo  {journal} {Phys. Rev. Lett.}\ }\textbf {\bibinfo
  {volume} {123}},\ \bibinfo {pages} {180604} (\bibinfo {year}
  {2019})}\BibitemShut {NoStop}%
\bibitem [{\citenamefont {Seah}\ \emph {et~al.}(2019)\citenamefont {Seah},
  \citenamefont {Nimmrichter},\ and\ \citenamefont {Scarani}}]{SeahPRE2019}%
  \BibitemOpen
  \bibfield  {author} {\bibinfo {author} {\bibfnamefont {S.}~\bibnamefont
  {Seah}}, \bibinfo {author} {\bibfnamefont {S.}~\bibnamefont {Nimmrichter}}, \
  and\ \bibinfo {author} {\bibfnamefont {V.}~\bibnamefont {Scarani}},\ }\href
  {\doibase 10.1103/PhysRevE.99.042103} {\bibfield  {journal} {\bibinfo
  {journal} {Phys. Rev. E}\ }\textbf {\bibinfo {volume} {99}},\ \bibinfo
  {pages} {042103} (\bibinfo {year} {2019})}\BibitemShut {NoStop}%
\bibitem [{\citenamefont {\c{C}akmak}\ \emph {et~al.}(2019)\citenamefont
  {\c{C}akmak}, \citenamefont {Campbell}, \citenamefont {Vacchini},
  \citenamefont {M\"ustecapl{\i}o\u{g}lu},\ and\ \citenamefont
  {Paternostro}}]{CakmakPRA2019}%
  \BibitemOpen
  \bibfield  {author} {\bibinfo {author} {\bibfnamefont {B.}~\bibnamefont
  {\c{C}akmak}}, \bibinfo {author} {\bibfnamefont {S.}~\bibnamefont
  {Campbell}}, \bibinfo {author} {\bibfnamefont {B.}~\bibnamefont {Vacchini}},
  \bibinfo {author} {\bibfnamefont {{\"O}.~E.}\ \bibnamefont
  {M\"ustecapl{\i}o\u{g}lu}}, \ and\ \bibinfo {author} {\bibfnamefont
  {M.}~\bibnamefont {Paternostro}},\ }\href {\doibase
  10.1103/PhysRevA.99.012319} {\bibfield  {journal} {\bibinfo  {journal} {Phys.
  Rev. A}\ }\textbf {\bibinfo {volume} {99}},\ \bibinfo {pages} {012319}
  (\bibinfo {year} {2019})}\BibitemShut {NoStop}%
\bibitem [{\citenamefont {Guarnieri}\ \emph {et~al.}(2020)\citenamefont
  {Guarnieri}, \citenamefont {Morrone}, \citenamefont {\c{C}akmak},
  \citenamefont {Plastina},\ and\ \citenamefont {Campbell}}]{GuarnieriPLA2020}%
  \BibitemOpen
  \bibfield  {author} {\bibinfo {author} {\bibfnamefont {G.}~\bibnamefont
  {Guarnieri}}, \bibinfo {author} {\bibfnamefont {D.}~\bibnamefont {Morrone}},
  \bibinfo {author} {\bibfnamefont {B.}~\bibnamefont {\c{C}akmak}}, \bibinfo
  {author} {\bibfnamefont {F.}~\bibnamefont {Plastina}}, \ and\ \bibinfo
  {author} {\bibfnamefont {S.}~\bibnamefont {Campbell}},\ }\href {\doibase
  https://doi.org/10.1016/j.physleta.2020.126576} {\bibfield  {journal}
  {\bibinfo  {journal} {Physics Letters A}\ }\textbf {\bibinfo {volume}
  {384}},\ \bibinfo {pages} {126576} (\bibinfo {year} {2020})}\BibitemShut
  {NoStop}%
\bibitem [{\citenamefont {Garc{\'\i}a-P{\'e}rez}\ \emph
  {et~al.}(2020)\citenamefont {Garc{\'\i}a-P{\'e}rez}, \citenamefont {Rossi},\
  and\ \citenamefont {Maniscalco}}]{garcia2020ibm}%
  \BibitemOpen
  \bibfield  {author} {\bibinfo {author} {\bibfnamefont {G.}~\bibnamefont
  {Garc{\'\i}a-P{\'e}rez}}, \bibinfo {author} {\bibfnamefont {M.~A.}\
  \bibnamefont {Rossi}}, \ and\ \bibinfo {author} {\bibfnamefont
  {S.}~\bibnamefont {Maniscalco}},\ }\href@noop {} {\bibfield  {journal}
  {\bibinfo  {journal} {npj Quantum Information}\ }\textbf {\bibinfo {volume}
  {6}},\ \bibinfo {pages} {1} (\bibinfo {year} {2020})}\BibitemShut {NoStop}%
\bibitem [{\citenamefont {Li}\ \emph {et~al.}(2020)\citenamefont {Li},
  \citenamefont {Chen}, \citenamefont {Xia}, \citenamefont {Zhang},\ and\
  \citenamefont {Man}}]{Li_2020}%
  \BibitemOpen
  \bibfield  {author} {\bibinfo {author} {\bibfnamefont {X.-M.}\ \bibnamefont
  {Li}}, \bibinfo {author} {\bibfnamefont {Y.-X.}\ \bibnamefont {Chen}},
  \bibinfo {author} {\bibfnamefont {Y.-J.}\ \bibnamefont {Xia}}, \bibinfo
  {author} {\bibfnamefont {Q.}~\bibnamefont {Zhang}}, \ and\ \bibinfo {author}
  {\bibfnamefont {Z.-X.}\ \bibnamefont {Man}},\ }\href {\doibase
  10.1088/1674-1056/ab84d0} {\bibfield  {journal} {\bibinfo  {journal} {Chinese
  Physics B}\ }\textbf {\bibinfo {volume} {29}},\ \bibinfo {pages} {060302}
  (\bibinfo {year} {2020})}\BibitemShut {NoStop}%
\bibitem [{\citenamefont {Cilluffo}\ \emph {et~al.}(2020)\citenamefont
  {Cilluffo}, \citenamefont {Carollo}, \citenamefont {Lorenzo}, \citenamefont
  {Gross}, \citenamefont {Palma},\ and\ \citenamefont
  {Ciccarello}}]{cilluffo2020collisional}%
  \BibitemOpen
  \bibfield  {author} {\bibinfo {author} {\bibfnamefont {D.}~\bibnamefont
  {Cilluffo}}, \bibinfo {author} {\bibfnamefont {A.}~\bibnamefont {Carollo}},
  \bibinfo {author} {\bibfnamefont {S.}~\bibnamefont {Lorenzo}}, \bibinfo
  {author} {\bibfnamefont {J.~A.}\ \bibnamefont {Gross}}, \bibinfo {author}
  {\bibfnamefont {G.~M.}\ \bibnamefont {Palma}}, \ and\ \bibinfo {author}
  {\bibfnamefont {F.}~\bibnamefont {Ciccarello}},\ }\href@noop {} {\enquote
  {\bibinfo {title} {Collisional picture of quantum optics with giant
  emitters},}\ } (\bibinfo {year} {2020}),\ \Eprint
  {http://arxiv.org/abs/2006.08631} {arXiv:2006.08631 [quant-ph]} \BibitemShut
  {NoStop}%
\bibitem [{\citenamefont {{De Chiara}}\ \emph {et~al.}(2018)\citenamefont {{De
  Chiara}}, \citenamefont {Landi}, \citenamefont {Hewgill}, \citenamefont
  {Reid}, \citenamefont {Ferraro}, \citenamefont {Roncaglia},\ and\
  \citenamefont {Antezza}}]{DeChiaraNJP2018}%
  \BibitemOpen
  \bibfield  {author} {\bibinfo {author} {\bibfnamefont {G.}~\bibnamefont {{De
  Chiara}}}, \bibinfo {author} {\bibfnamefont {G.}~\bibnamefont {Landi}},
  \bibinfo {author} {\bibfnamefont {A.}~\bibnamefont {Hewgill}}, \bibinfo
  {author} {\bibfnamefont {B.}~\bibnamefont {Reid}}, \bibinfo {author}
  {\bibfnamefont {A.}~\bibnamefont {Ferraro}}, \bibinfo {author} {\bibfnamefont
  {A.~J.}\ \bibnamefont {Roncaglia}}, \ and\ \bibinfo {author} {\bibfnamefont
  {M.}~\bibnamefont {Antezza}},\ }\href {\doibase 10.1088/1367-2630/aaecee}
  {\bibfield  {journal} {\bibinfo  {journal} {New Journal of Physics}\ }\textbf
  {\bibinfo {volume} {20}},\ \bibinfo {pages} {113024} (\bibinfo {year}
  {2018})}\BibitemShut {NoStop}%
\bibitem [{\citenamefont {Daryanoosh}\ \emph {et~al.}(2018)\citenamefont
  {Daryanoosh}, \citenamefont {Baragiola}, \citenamefont {Guff},\ and\
  \citenamefont {Gilchrist}}]{DaryanooshPRA2018}%
  \BibitemOpen
  \bibfield  {author} {\bibinfo {author} {\bibfnamefont {S.}~\bibnamefont
  {Daryanoosh}}, \bibinfo {author} {\bibfnamefont {B.~Q.}\ \bibnamefont
  {Baragiola}}, \bibinfo {author} {\bibfnamefont {T.}~\bibnamefont {Guff}}, \
  and\ \bibinfo {author} {\bibfnamefont {A.}~\bibnamefont {Gilchrist}},\ }\href
  {\doibase 10.1103/PhysRevA.98.062104} {\bibfield  {journal} {\bibinfo
  {journal} {Phys. Rev. A}\ }\textbf {\bibinfo {volume} {98}},\ \bibinfo
  {pages} {062104} (\bibinfo {year} {2018})}\BibitemShut {NoStop}%
\bibitem [{\citenamefont {Esposito}\ \emph {et~al.}(2010)\citenamefont
  {Esposito}, \citenamefont {Lindenberg},\ and\ \citenamefont {den
  Broeck}}]{EspositoNJP2010}%
  \BibitemOpen
  \bibfield  {author} {\bibinfo {author} {\bibfnamefont {M.}~\bibnamefont
  {Esposito}}, \bibinfo {author} {\bibfnamefont {K.}~\bibnamefont
  {Lindenberg}}, \ and\ \bibinfo {author} {\bibfnamefont {C.~V.}\ \bibnamefont
  {den Broeck}},\ }\href {\doibase 10.1088/1367-2630/12/1/013013} {\bibfield
  {journal} {\bibinfo  {journal} {New Journal of Physics}\ }\textbf {\bibinfo
  {volume} {12}},\ \bibinfo {pages} {013013} (\bibinfo {year}
  {2010})}\BibitemShut {NoStop}%
\bibitem [{\citenamefont {Reeb}\ and\ \citenamefont
  {Wolf}(2014)}]{ReebNJP2014}%
  \BibitemOpen
  \bibfield  {author} {\bibinfo {author} {\bibfnamefont {D.}~\bibnamefont
  {Reeb}}\ and\ \bibinfo {author} {\bibfnamefont {M.~M.}\ \bibnamefont
  {Wolf}},\ }\href {\doibase 10.1088/1367-2630/16/10/103011} {\bibfield
  {journal} {\bibinfo  {journal} {New Journal of Physics}\ }\textbf {\bibinfo
  {volume} {16}},\ \bibinfo {pages} {103011} (\bibinfo {year}
  {2014})}\BibitemShut {NoStop}%
\bibitem [{\citenamefont {Ptaszy\ifmmode~\acute{n}\else \'{n}\fi{}ski}\ and\
  \citenamefont {Esposito}(2019)}]{PtaszynskiPRL2019}%
  \BibitemOpen
  \bibfield  {author} {\bibinfo {author} {\bibfnamefont {K.}~\bibnamefont
  {Ptaszy\ifmmode~\acute{n}\else \'{n}\fi{}ski}}\ and\ \bibinfo {author}
  {\bibfnamefont {M.}~\bibnamefont {Esposito}},\ }\href {\doibase
  10.1103/PhysRevLett.123.200603} {\bibfield  {journal} {\bibinfo  {journal}
  {Phys. Rev. Lett.}\ }\textbf {\bibinfo {volume} {123}},\ \bibinfo {pages}
  {200603} (\bibinfo {year} {2019})}\BibitemShut {NoStop}%
\bibitem [{\citenamefont {Campisi}\ \emph {et~al.}(2015)\citenamefont
  {Campisi}, \citenamefont {Pekola},\ and\ \citenamefont
  {Fazio}}]{CampisiNJP2015}%
  \BibitemOpen
  \bibfield  {author} {\bibinfo {author} {\bibfnamefont {M.}~\bibnamefont
  {Campisi}}, \bibinfo {author} {\bibfnamefont {J.}~\bibnamefont {Pekola}}, \
  and\ \bibinfo {author} {\bibfnamefont {R.}~\bibnamefont {Fazio}},\ }\href
  {\doibase 10.1088/1367-2630/17/3/035012} {\bibfield  {journal} {\bibinfo
  {journal} {New Journal of Physics}\ }\textbf {\bibinfo {volume} {17}},\
  \bibinfo {pages} {035012} (\bibinfo {year} {2015})}\BibitemShut {NoStop}%
\bibitem [{\citenamefont {Ozols}(2009)}]{Ozols}%
  \BibitemOpen
  \bibfield  {author} {\bibinfo {author} {\bibfnamefont {M.}~\bibnamefont
  {Ozols}},\ }\href
  {http://home.lu.lv/~sd20008/papers/essays/Random%20unitary%20[paper].pdf}
  {\enquote {\bibinfo {title} {How to generate a random unitary matrix},}\ }
  (\bibinfo {year} {2009})\BibitemShut {NoStop}%
\bibitem [{\citenamefont {Ollivier}\ and\ \citenamefont
  {Zurek}(2001)}]{PhysRevLett.88.017901}%
  \BibitemOpen
  \bibfield  {author} {\bibinfo {author} {\bibfnamefont {H.}~\bibnamefont
  {Ollivier}}\ and\ \bibinfo {author} {\bibfnamefont {W.~H.}\ \bibnamefont
  {Zurek}},\ }\href {\doibase 10.1103/PhysRevLett.88.017901} {\bibfield
  {journal} {\bibinfo  {journal} {Phys. Rev. Lett.}\ }\textbf {\bibinfo
  {volume} {88}},\ \bibinfo {pages} {017901} (\bibinfo {year}
  {2001})}\BibitemShut {NoStop}%
\bibitem [{\citenamefont {Henderson}\ and\ \citenamefont
  {Vedral}(2001)}]{henderson2001classical}%
  \BibitemOpen
  \bibfield  {author} {\bibinfo {author} {\bibfnamefont {L.}~\bibnamefont
  {Henderson}}\ and\ \bibinfo {author} {\bibfnamefont {V.}~\bibnamefont
  {Vedral}},\ }\href
  {http://iopscience.iop.org/article/10.1088/0305-4470/34/35/315/meta}
  {\bibfield  {journal} {\bibinfo  {journal} {Journal of physics A:
  mathematical and general}\ }\textbf {\bibinfo {volume} {34}},\ \bibinfo
  {pages} {6899} (\bibinfo {year} {2001})}\BibitemShut {NoStop}%
\bibitem [{\citenamefont {{De Chiara}}\ and\ \citenamefont
  {Antezza}(2020)}]{datarep}%
  \BibitemOpen
  \bibfield  {author} {\bibinfo {author} {\bibfnamefont {G.}~\bibnamefont {{De
  Chiara}}}\ and\ \bibinfo {author} {\bibfnamefont {M.}~\bibnamefont
  {Antezza}},\ }\href {\doibase 10.17034/46387195-ff10-4b6c-8725-a1fed5bad759}
  {\enquote {\bibinfo {title} {{Dataset} for all the plots in the paper.}}\ }
  (\bibinfo {year} {2020})\BibitemShut {NoStop}%
\end{thebibliography}%
%

\end{document}